\documentclass[showpacs,preprintnumbers,amsmath,amssymb,superscriptaddress]{revtex4-1}
\usepackage{ulem}
\usepackage[dvips]{graphicx}
\usepackage{color}
\usepackage{bm}
\numberwithin{equation}{section}
\begin{document}
\title{Calculable microscopic theory for $^{12}$C($\alpha, \gamma$)$^{16}$O cross section near Gamow window}
\author{Y. Suzuki}
\affiliation{Department of Physics, Niigata University, Niigata 950-2181, Japan}
\affiliation{RIKEN Nishina Center, Wako 351-0198, Japan}
\email{yasuyuki$_$suzuki@riken.jp}
\date{\today}

\begin{abstract}
\noindent
{\bf Abstract} $^{12}$C$(\alpha, \gamma)^{16}$O radiative-capture process is a key reaction to produce the element of oxygen in stars. Measuring the cross section near the Gamow window 
is extremely hard because it is too small. To make a theoretical contribution 
towards resolving the long-standing problem, I present a microscopic formulation that aims at providing all materials needed to calculate the cross section.  
The  states of 
$^{12}$C and $^{16}$O relevant to the reaction are respectively described with fully microscopic 3 $\alpha$-particle and 4 $\alpha$-particle configurations, 
in which the relative motion among the $\alpha$ particles is expanded 
in terms of correlated Gaussian basis functions. The configuration space 
has the advantage 
of being able to well describe the reduced $\alpha$-width amplitudes of 
the states of $^{16}$O. Both electric dipole and electric quadrupole 
transitions are responsible for the radiative-capture process. 
The $\alpha$ particle is described with 
a $(0s)^4$ configuration admixed with a small amount of an isospin $T=1$ 
impurity component, which 
is crucially important to account for the isovector electric dipole transition. 
The isoscalar electric dipole operators are also taken into account up to the first 
order beyond the long-wavelength approximation. 
All the necessary ingredients are provided to make the paper self-contained and 
ready for numerical computations. 
\end{abstract}
\maketitle

\section{Introduction}

\subsection{Background}

The abundance of chemical elements in the universe is dominated by hydrogen and 
helium that 
were produced in the Big Bang. The two elements occupy almost 99.9 \% when the 
abundance is 
measured by the numerical count of atoms.  Among the rest of elements, oxygen, 
carbon, and neon follow. The synthesis of carbon is made via the 
triple-alpha reaction, and its abundance is accounted for by the existence of 
a narrow resonance 
state of $^{12}$C, the Hoyle state~\cite{hoyle54,cook57}. 

The abundance ratio of carbon to oxygen is about 0.6 by the numerical count. 
This ratio depends on how fast 
$^{12}$C($\alpha, \gamma$)$^{16}$O radiative-capture reaction occurs during helium-burning processes in stars. 
See Ref.~\cite{deboer17} for a recent review on the capture reaction 
as well as relevant references. The Maxwellian-averaged reaction rate is a key 
quantity to determine the 
nucleosynthesis of elements in stars. The rate for two-body reactions is 
computed from~\cite{nacre}
\begin{align}
\langle \sigma v \rangle=\Big(\frac{8}{\pi \mu}\Big)^{\frac{1}{2}}\frac{1}{(k_BT)^{\frac{3}{2}}} \int_0^{\infty}dE \, \sigma(E) E\, e^{-\frac{E}{k_BT}},
\end{align}
where $\mu$ the reduced mass, $k_B$ the Boltzmann constant, 
$T$ the temperature, $\sigma(E)$ the cross section at the energy $E$ in the center-of-mass (c.m.) system, $v$ the relative velocity. 
The cross section can be expressed in terms of an astrophysical $S$-factor, 
$S(E)$, as 
\begin{align}
\sigma(E)E=S(E)e^{-2\pi \eta},
\end{align}
where $\eta$ is the Sommerfeld parameter, $\eta=\frac{Z_1Z_2e^2}{\hbar v}$.

The Gamow peak of 
the $^{12}$C($\alpha, \gamma$)$^{16}$O reaction is about $E_0=0.3$ MeV. 
If there is a resonance in the reaction 
near the Gamow window corresponding to helium-burning 
temperatures, $0.1-0.2\, T_9$, the reaction proceeds quickly. 
However, there is no resonance in $^{16}$O up to $2\,T_9$, which is in 
contrast to the case of $^{12}$C. 
See, e.g., Fig. 7.5 of Ref.~\cite{rolfs88}. Thus one must consider other 
mechanism 
to account for the abundance ratio. Two processes are considered: One is a 
nonresonant direct capture (DC) process, and the other a nonresonant type 
of capture into the tails of nearby states. The former includes a broad $1^-$ 
resonance (420 keV width) at $9.585$ MeV and the latter includes $6.917; 2^+$ 
and $7.117; 1^-$ states~\cite{nudat}. The astrophysical $S$-factors at $E_0$ 
are estimated, in units of {\rm MeV\,b}, to be~\cite{rolfs88}
\begin{align}
S_{\rm DC}(E_0)\approx 5\times 10^{-3},\ \ \ \ \ S_{9.585}(E_0)\approx 1.5\times 10^{-3},\ \ \ \ \ S_{6.917}(E_0)\approx 0.2,\ \ \ \ \ S_{7.117}(E_0)\approx 0.1.
\end{align}
Both the DC and the low-energy tail of the 9.585 MeV resonance 
are relatively insignificant compared with those of the two subthreshold 
states. It appears that the $^{12}$C($\alpha, \gamma$)$^{16}$O 
burning rate is determined almost entirely by the effects of these 
subthreshold states: $S(E_0)\approx 0.3$ MeV$\,$b.
Important ingredients are then the reduced $\alpha$-widths of those 
states and the electric quadrupole ($E2$)  and electric dipole ($E1$) transition probabilities to the ground state. 
The astrophysical $S$-factor of  
$^{12}$C$(\alpha, \gamma)^{16}$O reaction determined phenomenologically~\cite{nacre} is known as a standard value. See Ref.~\cite{katsuma08a,katsuma08b,katsuma08c} for a potential model study of the radiative-capture reaction as well as Ref.~\cite{descouvemont10} for its comment. 

One has to calculate both the $E1$ and $E2$ transitions from 
the $\alpha+^{12}$C 
continuum states to the ground state of $^{16}$O. The $E1$ transition in 
$^{16}$O is isospin-forbidden in the zeroth-order because both 
the $1^-$ state and the ground state have isospin $T=0$ almost perfectly. 
One must 
consider possible mechanism to take account of the $E1$ transition. I take 
up two possibilities:  
One is to include the isospin mixing in $^{16}$O, and another is to take 
into account the next-order term, isoscalar (IS) $E1$ operators. 

The isospin mixing is primarily induced by the Coulomb interaction. 
The isospin mixing in $^{16}$O was treated 
by taking into account, e.g., the coupling to the $^{15}$N$+p$ and $^{15}$O$+n$ 
channels~\cite{pdesc87,pdesc87b}. 
It is very unlikely, however, that either $^{15}$N or $^{15}$O is present in the process of 
$^{12}$C($\alpha, \gamma$)$^{16}$O reaction when the stars are still at 
too low temperatues to ignite the CNO cycle. Looking for other 
possible mechanism for the isospin mixing appears to be natural. 
A possible mechanism  is 
the distortion of $\alpha$ particle itself from a pure $T=0$ state: 
\begin{align}
\phi_{\alpha}=\sqrt{1-\epsilon^2}\phi_{\alpha}^{(0)}+\epsilon \phi_{\alpha}^{(1)},
\label{alpha*}
\end{align}
where $\phi_{\alpha}^{(0)}$ is a dominant wave function of $\alpha$ particle 
with $T=0$, and $\phi_{\alpha}^{(1)}$ stands for a $T=1$ impurity state. They 
both have parity $\pi=+$ and are assumed to have orbital angular momentum  
$L=0$ and spin $S=0$. 
$|\epsilon|$ is on the order of $4\times 10^{-3}$~\cite{wiringa} although 
it depends on a nuclear force employed. 
I discuss how to construct $\phi^{(1)}_{\alpha}$ in Sec.~\ref{isf.alpha}. 
Since all the states in $^{12}$C and $^{16}$O that are relevant to the 
radiative-capture reaction are expected to be 
described fairly well by $3\, \alpha$- and $4\, \alpha$-particle 
configurations, 
taking account of the distortion of $\alpha$ particle immediately 
brings about the 
isospin mixing in $^{12}$C and $^{16}$O as well. 

One of the IS $E1$ operators is given by  
$-\frac{k^2}{10}r^3Y_{1\mu}(\hat{\bm r})$ if the spin and orbital 
angular-momentum dependent term is omitted~\cite{baye12}. Compared to 
the isovector (IV) $E1$ operator of leading-order, $rY_{1\mu}(\hat{\bm r})$, 
the contribution of the IS $E1$ operator is estimated to be larger by $f=2\frac{k^2}{10}r^2$, where the factor 2 
takes account of the IS nature of the operator. 
For the $E1$ transition from the 7.117;1$^-$ state to the ground state of 
$^{16}$O, the wave number $k$ is 0.036\,fm$^{-1}$. If $r$ is taken to be 
2.5\, fm, about the radius of $^{16}$O, $f \approx 2\times 10^{-3}$, which 
is comparable to $|\epsilon|$. The above estimate suggests that it is safe to take into account both IV and IS $E1$ operators. 
Refer to Ref.~\cite{baye12} for discussions on 
the IS $E1$ operators of the 
first-order beyond the long-wavelength approximation.

\subsection{Conditions of calculation}
\label{condition}

The guideline of calculation is to use a microscopic model, 
reproduce the energies (measured from the $\alpha$ threshold of $^{16}$O) of 
the states relevant to the radiative-capture reaction, determine effective charges for $p-$ and $d-$wave radiative captures by reproducing the 
$E1$ and 
$E2$ transition rates of the relevant subthreshold states, and improve the reduced $\alpha$-width amplitudes or the asymptotic normalization constants  
as much as possible. After this preparation one can get 
the radiative-capture cross section reliably in the framework of a microscopic 
$R$-matrix theory~\cite{desc10}. An example of the present study is 
the calculation 
of the astrophysical $S$-factors of four-nucleon system~\cite{arai11}. 

It is meant by the microscopic model that the antisymmetrization requirement 
on the constituent nucleons is fully taken into account. The condition is particularly 
important in the synthesis of $^{16}$O even though the $4\, \alpha$-particle 
model is employed, because the nuclear shell structure manifests itself 
clearly in $^{16}$O. Note, however, that the configuration space spanned by the microscopic cluster model has significant overlap with low-lying shell-model configuration space. See, e.g., Refs.~\cite{brink66,horiuchi77,suzuki76a,suzuki76b,suzuki96}. In this respect the synthesis of $^{16}$O is in a 
contrast to that of $^{12}$C by the 
triple-alpha reaction, where a boson model of $3\, \alpha$-particles~\cite{ishikawa13,suno16} presents the reaction rate consistently with 
Ref.~\cite{nacre}. 

One has to get the wave functions of the states with $L^{\pi}=0^+_1, 0^+_2, 
2^+_1, 1^-_1$, (and $1^-_2$). ($L$ is used instead of $J$ to denote 
the total angular momentum because the total spin is restricted to zero.) 
These states are described by a microscopic $4\alpha$ model combined with a 
$^{12}$C$(0^+)+\alpha$ model. The latter component is important 
to properly describe  the tail of the reduced $\alpha$-width 
amplitude. The $\alpha$ 
particle in the microscopic calculation is represented by $\phi_{\alpha}$ of 
Eq.~(\ref{alpha*}), but it is replaced by $\phi^{(0)}_{\alpha}$ in actual calculations, and the wave 
functions of the relevant states are all determined in the usual 4$\alpha$ 
cluster model, namely under $\epsilon=0$. This treatment 
appears reasonable because $\epsilon$ is so small that a  
possible change in both the energies and properties of the relevant states is 
expected to be negligibly small except for the $E1$ transition rate. 
The wave functions of all the states are thus determined by 
assuming $\epsilon=0$, and 
only when calculating the isospin-forbidden or IV $E1$ transition rate, 
one of $\phi_{\alpha}^{(0)}$'s of the 
wave functions is replaced by $\epsilon \phi_{\alpha}^{(1)}$. 

One may change the parameters of a nucleon-nucleon interaction depending on 
$L^{\pi}$ in order to reproduce precisely 
the energies from $^{12}$C$+\alpha$ threshold. 
Remember that the $1_1^-$ state is located only 45 keV lower than the $\alpha$ 
threshold. Obviously the reduced $\alpha$-width of 
that state should play an important role in the $p$-wave radiative 
capture. The $\beta$-delayed $\alpha$-spectrum of $^{16}$N 
constrains its reduced $\alpha$-width~\cite{baye88,azuma94,tang10}, estimating 
$S_{7.117}(E_0)\approx 86\,\pm \,22$\, keV\,b~\cite{tang10}.

The conditions to be met in the calculation are the following:

\vspace{1mm}
\noindent
(1) Calculate the ground-state energy of $^{12}$C in the 3$\alpha$ model.
The $^{12}$C+$\alpha$ threshold is defined by that value. The ground-state 
wave function of $^{12}$C is used to describe $^{12}$C+$\alpha$ configurations 
in the initial channel. 

\vspace{1mm}
\noindent
(2) Adjust the potential parameter to reproduce $E(0^+_2)$, the energy of the second $0^+$ state of $^{16}$O, which is 1.11\,MeV lower than  
the $^{12}$C+$\alpha$ threshold. Fulfill conditions (1) and (2) 
simultaneously. No problem that 
$E(0^+_1)$ is not reproduced very well, but its root-mean-square radius should be reasonably well reproduced. 

\vspace{1mm} 
\noindent
(3) Adjust the potential parameters for $1^-$ and $2^+$ to reproduce $E(1^-_1)$ and $E(2^+_1)$ from 
the $^{12}$C+$\alpha$ threshold, respectively. It is interesting to know where 
$E(1^-_2)$ and $E(2^+_2)$ are predicted. 

\vspace{1mm}
\noindent
(4) Determine an $E2$ effective charge so as to reproduce the experimental 
$E2$ transition rate for $2^+_1 \to 0^+_1$. With that effective charge 
calculate the cross section for the $d$-wave radiative capture of 
$\alpha$ particle to the $0^+_1$ state. 

\vspace{1mm}
\noindent
(5) Calculate the $E1$ transition rate for $1^-_1 \to 0^+_1$ as explained 
below. Determine an $E1$ effective charge or adjust $\epsilon$ to reproduce the 
experimental value. Calculate the $p$-wave radiative-capture 
cross section of $\alpha$ particle to the $0^+_1$ state. 

\vspace{1mm}
\noindent
There is hope that the above requirements (1) to (5) are all met in 
view of the achievement of semi-microscopic~\cite{suzuki76a,suzuki76b} and microscopic~\cite{pdesc87} $^{12}$C$+\alpha$ studies. 

The $E2$ and $E1$ transitions indicated above are both  
one-step processes from the continuum state to the ground state. 
Another type of 
possible transition is a two-step (or cascade) process from the $s$-wave continuum state to the ground state via the $2_1^+$ state for $E2$ or the $1_1^-$ state 
for $E1$. Although the two-step process is favored by the
absence of the centrifugal barrier, its transition rate is considerably small  
compared to the one-step process~\cite{pdesc87b}.

The leading $E1$ operator is IV and its next-order term is IS: $E1=E1({\rm IV})+E1({\rm IS})$. Once the relevant wave functions are obtained, the $E1$ matrix element reads as 
$\langle \Psi^{0^+0}|E1|\Psi^{1^-M}\rangle=A_{{\rm IV}}+A_{{\rm IS}}$,
where the isospin-forbidden $E1$ matrix element, $A_{{\rm IV}}=\big\langle \Psi^{0^+0}(\phi(4\alpha))|E1(\rm IV)|\Psi^{1^-M}(\phi(4\alpha))\big\rangle$, is calculated from 
\begin{align}
A_{{\rm IV}}
\to \epsilon   \sum_{i=1}^4 \Big[
 \big\langle \Psi^{0^+0}(i)|E1({\rm IV})|\Psi^{1^-M}(\phi^{(0)}(4\alpha))\big\rangle  +  \big\langle \Psi^{0^+0}(\phi^{(0)}(4\alpha))|E1({\rm IV})|\Psi^{1^-M}(i) \big\rangle \Big],
\label{E1mat.ele}
\end{align} 
where $\phi(4\alpha)=\phi_{\alpha}(1)\cdots \phi_{\alpha}(4)$, 
$\phi^{(0)}(4\alpha)=\phi^{(0)}_{\alpha}(1)\cdots \phi^{(0)}_{\alpha}(4)$, and 
$\Psi^{0^+0}(i)$ is the wave function defined by replacing 
$\phi_{\alpha}(i)$ in $\Psi^{0^+0}(\phi(4\alpha))$ by $\phi_{\alpha}^{(1)}(i)$ and the remaining ones by $\phi^{(0)}_{\alpha}$'s. 
$\Psi^{1^-M}(i)$ is defined from $\Psi^{1^-M}(\phi(4\alpha))$ similarly. See Sec.~\ref{cal.isfE1} for details. The ansatz is reasonable because the leading term $\langle \Psi^{0^+0}(\phi^{(0)}(4\alpha))|E1({\rm IV})|\Psi^{1^-M}(\phi^{(0)}(4\alpha))\rangle$ vanishes and $\epsilon$ is very small.  On the other hand, the IS $E1$ matrix element, $A_{\rm IS}=\langle \Psi^{0^+0}(\phi(4\alpha))|E1({\rm IS})|\Psi^{1^-M}(\phi(4\alpha))\rangle$, 
and the $E2$ transition rate are calculated from the main 
components of the wave functions as follows:
\begin{align}
&A_{\rm IS}
\to \langle \Psi^{0^+0}(\phi^{(0)}(4\alpha))|E1({\rm IS})|\Psi^{1^-M}(\phi^{(0)}(4\alpha))\rangle,\\
\label{ISE1mat}
&\langle \Psi^{0^+0}(\phi(4\alpha))|E2|\Psi^{2^+M}(\phi(4\alpha))\rangle 
\to 
\langle \Psi^{0^+0}(\phi^{(0)}(4\alpha))|E2|\Psi^{2^+M}(\phi^{(0)}(4\alpha))\rangle.
\end{align}

\subsection{Outline of basis functions}
\label{outline.basis}

I present a framework to study the radiative capture of $\alpha$ particle by $^{12}$C at low energies using a 
microscopic $4\alpha$ cluster model. The four $\alpha$-particles with a 
good angular momentum describe not only the relevant bound states of $^{16}$O 
but also $^{12}$C$+\alpha$ continuum states. In the latter case $^{12}$C 
decribed with the $3\alpha$ cluster model has a good 
angular momentum and the relative motion between $^{12}$C and $\alpha$-particle 
also has a good orbital angular momentum. The radiative-capture 
problem demands calculations allowing for a variety of $4\alpha$ structure 
as well as double angular-momentum projections. To cope with these requirements 
I employ a correlated Gaussian (CG) basis with global-vector representation, 
following the formulation of Refs.~\cite{varga95,book98}. 

The antisymmetrized CG basis for $N\alpha$-particle system is defined by
\begin{align}
\Psi_{K}^{LM}(u,A)=\frac{1}{{\sqrt{4!}}^N}{\cal A}_{4N}\left\{f_{KLM}(u,A,\bm x) \phi^{\rm in}(N\alpha)\right\}, 
\label{CGbasis}
\end{align}
where ${\cal A}_{4N}$ is an antisymmetrizer, 
${\cal A}_{4N}=\frac{1}{\sqrt{(4N)!}}\sum_p \epsilon(p)p$, summed over $(4N)!$ 
permutations $p$ with their phases $\epsilon(p)$. It satisfies ${\cal A}_{4N}^2=\sqrt{(4N)!}{\cal A}_{4N}$. 
$\phi^{\rm in}(N\alpha)$ stands for the product of the internal wave functions 
of $N\, \alpha$-particles,
\begin{align}
\phi^{\rm in}(N\alpha)=\prod_{i=1}^N \phi_{\alpha}(i) \to \prod_{i=1}^N \phi_{\alpha}^{(0)}(i),
\label{prod.alpha.internal}
\end{align} 
where both $\phi_{\alpha}(i)$ and $\phi_{\alpha}^{(0)}(i)$ are normalized, antisymmetrized and contain no 
c.m. motion of the $\alpha$ particle. The basis~(\ref{CGbasis}) is 
characterized by $f_{KLM}(u,A,\bm x)$, a prototype of the CG basis with 
the global-vector representation, which here describes the motion 
among $\alpha$ particles. It is defined by 
\begin{align}
&f_{KLM}(u,A,\bm x)=|\tilde{u}\bm x|^{2K}{\cal Y}_{LM}(\tilde{u}\bm x) \, 
e^{-\frac{1}{2}\tilde{\bm x}A\bm x},\ \ \ \ \ {\cal Y}_{LM}(\bm r)=r^{L}Y_{LM}(\hat{\bm r}).
\label{def.fLM}
\end{align} 
The coordinate $\bm x$ is a column vector of $N-1$ dimension comprising 
relative distance vectors among $\alpha$ particles. Its definition 
is given in  Sec.~\ref{matele.cg}. $K$ is a non-negative integer parameter 
mostly set to 0, but one may have to extend it to small integers 
when the energy gain becomes significantly large and/or the electric transition rates are 
better reproduced. Both $A$ and $u$ are variational parameters:  
$A$ is an  $(N-1)\times (N-1)$ positive-definite symmetric matrix and  $u$ is 
a column vector of  $N-1$ dimension.  The tilde symbol $\,\tilde{\,\,}\,$ indicates the transpose of 
a matrix or a column vector: E.g., the global vector $\tilde{u}\bm x$ in Eq.~(\ref{def.fLM}) stands 
for  $\tilde{u}\bm x\equiv \sum_{i=1}^{N-1}u_i{\bm x}_i$. The dot for a scalar product of 
two three-dimensional vectors is often omitted: E.g., $\tilde{\bm x}A\bm x$ stands for $\sum_{i=1}^{N-1}{\bm x}_i \cdot (A\bm x)_i
=\sum_{i,j=1}^{N-1}A_{ij}{\bm x}_i\cdot {\bm x}_j$. $u$ is assumed to be  normalized to $\tilde{u}u=1$ without  loss of generality.
The CG basis~(\ref{CGbasis}) has a good angular momentum and parity $\pi=(-1)^L$, and it  
is translation-invariant: The exponential $e^{-\frac{1}{2}\tilde{\bm x}A\bm x}$ is rotation-invariant, while  ${\cal Y}_{LM}(\tilde{u}\bm x)$ describes the rotational motion. $\hat{\bm r}$ stands for the polar angle and the azimuthal angle of $\bm r$. 

The total wave function $\Psi^{LM}$ is given by a 
combination of the CG basis functions,
\begin{align}
\Psi^{LM}=\sum_i C_i \Psi_{K_i}^{LM}(u_i, A_i).
\label{totalwf}
\end{align}
The parameters $K_i$, $u_i$, and $A_i$ may be chosen with, e.g., 
the stochastic variational method~\cite{varga94,varga95,book98} for 
bound states. 
For scattering states, they have to be chosen to account for the asymptotics 
of the scattering wave function. 
A matrix element between the CGs is obtained by an integral transform 
of the one between Slater determinants used in $\alpha$-cluster model. 
Although the latter matrix element is quite familiar in the nuclear cluster model~\cite{brink66,horiuchi77}, one has to obtain it 
{\it analytically} as a function of the positions of $\alpha$ clusters to make the transform practicable~\cite{varga95,book03}.  This kind of 
analytic calculation was used to study possible Bose-Einstein condensation in $^{12}$C with $3\alpha$ model~\cite{matsumura04}. 

One may need more global vectors, e.g.,  
$[{\cal Y}_{L_1}(\widetilde{u_1}\bm x)\times {\cal Y}_{L_2}(\widetilde{u_2}\bm x)]_{LM}$ to describe more complicated rotational motion, as discussed in Refs.~\cite{suzuki08,aoyama12}.
A suitable choice of $u$ and $A$ of the single global vector, however, makes it 
possible to represent 
certain types of doubly projected states. See Sec.~\ref{choice.cgparam} 
for the case 
of $L_1=0, \, L_2=L$, which is most important for the present purpose. 
In both calculations of Refs.~\cite{suzuki76a,suzuki76b,pdesc87}, the configuration of $^{12}$C(2$^+$)+$\alpha$ channel is explicitly included, and is found to 
play a vital role in reproducing some high-lying states of $^{16}$O.  
There are two reasons why the configuration is not included explicitly 
in this study. One is that the states of present interest are all 
low-lying states and expected to be described well without its explicit 
inclusion. Second is that the basis functions used 
here are fully correlated and can take account 
of $^{12}$C(2$^+$)+$\alpha$ configurations to some extent~\cite{book98,suzuki98}.  
If two global vectors are definitely needed, the procedure explained in 
Sec.~\ref{transf.slater.cg} has to be extended appropriately. 

This paper is organized as follows. 
Section~\ref{Nalpha.me.cm.free} derives the analytic form for some important 
matrix elements between the Slater determinants 
of $N\,\alpha$-particle model and shows how to eliminate the excitation of 
the total c.m. motion.  Section~\ref{isf.alpha} details the construction of 
the $T=1$ impurity component $\phi^{(1)}_{\alpha}$ of $\alpha$ particle,  
explains its inclusion  in the $N\,\alpha$-particle model calculation, 
and derives the matrix element for the $E1({\rm IV})$ operator. 
Section~\ref{matele.cg}  performs the basis conversion from the Slater determinant to the CG, gives general formulas for the CG matrix elements, 
and briefly discusses the choice of the CG parameters together 
with its relation to 
the calculation of the reduced $\alpha$-width amplitude. Section~\ref{summary} is a short summary. Since 
this paper focuses on providing all ingredients needed for a 
numerical calculation, references are not exhaustively cited but restricted 
to those closely related to technical points.

\section{$N\,\alpha$-particle model}

\label{Nalpha.me.cm.free}

\subsection{Slater determinants of Gaussian wave-packets}

Calculating matrix elements between Slater determinants of $N \alpha$-particle 
system is crucially important for the present purpose.
An $N \alpha$-paticle  Slater determinant is defined by
\begin{align}
\phi(\{\bm S\})&={\cal A}_{4N}\left\{ 
\phi_{\bm S_1}^{\beta} \chi_1 \cdots \phi_{\bm S_N}^{\beta} \chi_1 \ 
\phi_{\bm S_1}^{\beta} \chi_2 \cdots \phi_{\bm S_N}^{\beta} \chi_2  \ 
 \phi_{\bm S_1}^{\beta} \chi_3 \cdots \phi_{\bm S_N}^{\beta} \chi_3  \ 
\phi_{\bm S_1}^{\beta} \chi_4 \cdots \phi_{\bm S_N}^{\beta} \chi_4 \right\},
\label{Nalpha.SD}
\end{align}
where    
$\phi_{\bm s}^{\beta}$ is a Gaussian wave-packet (GWP) centered at $\bm s$
\begin{align}
\phi_{\bm s}^{\beta}(\bm r)=\left(\frac{\beta}{\pi}\right)^{\frac{3}{4}}e^{-\frac{\beta}{2}(\bm r-\bm s)^2},
\label{gwp}
\end{align}
and $\chi_1=n\!\uparrow$, $\chi_2=n\!\downarrow$, $\chi_3=p\!\uparrow$, $\chi_4=p\!\downarrow$ stand for four spin-isospin states of the nucleon. Both neutrons and protons with spin up and spin down occupy the GWP centered at each $\bm S_i$ ($i=1,\ldots, N)$. Since this occupancy is always assumed in the $\alpha$-cluster model, no spin-isospin label is put in $\phi(\{\bm S\})$. The total spin and isospin of $\phi(\{\bm S\})$ are of course zero. 
Equation~(\ref{Nalpha.SD}) is equivalent to the usual form of the 
$\alpha$-cluster wave function~\cite{brink66}:
\begin{align}
\phi(\{\bm S\})&={\cal A}_{4N}\Big\{ 
\prod_{i=1}^N\Big( \phi_{\bm S_i}^{\beta} \chi_1 \phi_{\bm S_i}^{\beta} \chi_2  
\phi_{\bm S_i}^{\beta} \chi_3 \phi_{\bm S_i}^{\beta} \chi_4 \Big)\Big\}.
\label{Nalpha.SD'}
\end{align}
Let ${\cal A}_4$ denote an antisymmetrizer of $1, {N+1}, {2N+1}, {3N+1}$. 
The $T=0$ main component of $\alpha$ particle, 
$\phi_{\alpha}^{(0)}$, is defined as an antisymmetrized product of the GWPs: 
\begin{align}
{\cal A}_4\left\{\phi_{\bm S_1}^{\beta}(\bm r_1) \chi_1(1)\phi_{\bm S_1}^{\beta}(\bm r_{N+1}) \chi_2(N+1)\phi_{\bm S_1}^{\beta}(\bm r_{2N+1}) \chi_3(2N+1)\phi_{\bm S_1}^{\beta}(\bm r_{3N+1}) \chi_4(3N+1)\right\}
=\phi_{\bm S_1}^{4\beta}(\bm R_1)\phi_{\alpha}^{(0)}(1),
\label{partial.antsymmetrizer}
\end{align}
where $\phi_{\bm S_1}^{4\beta}(\bm R_1)$, $\bm R_1=\frac{1}{4}\sum_{i=1}^4\bm r_{(i-1)N+1}$, describes the c.m. motion of $\alpha$ 
particle centered at $\bm S_1$.  As  confirmed easily, 
the left-hand side of Eq.~(\ref{partial.antsymmetrizer}) is normalized. 
$\phi_{\alpha}^{(0)}$ is nothing but the $(0s)^4$ configuration. 
By noting that 
${\cal A}_{4N}{\cal A}_4={\cal A}_4 {\cal A}_{4N}=\sqrt{4!}{\cal A}_{4N}$ 
and by repeating the above antisymmetrization procedure for $i, N+i, 2N+i, 3N+i$ ($i=2,\ldots, N$), 
the Slater determinant~(\ref{Nalpha.SD}) or~(\ref{Nalpha.SD'}) reduces to 
the usual form 
\begin{align}
\phi(\{\bm S\})=\frac{1}{{\sqrt{4!}}^N}{\cal A}_{4N}\left\{
\prod_{i=1}^N \phi_{\bm S_i}^{4\beta}(\bm R_i)\phi_{\alpha}^{(0)}(i)\right\}.
\label{Nalpha.WF}
\end{align}

The exponent of $\prod_{i=1}^N \phi_{\bm S_i}^{4\beta}(\bm R_i)$ is 
$-\frac{4\beta}{2}\sum_{i=1}^N(\bm R_i-\bm S_i)^2$. 
The total c.m. motion of $\phi(\{\bm S\})$ is then given by the GWP, 
$\phi^{4N\beta}_{\overline{\bm S}}(\bm R)$, where $\bm R=\frac{1}{4N} \sum_{i=1}^{4N}\bm r_i=\frac{1}{N}\sum_{i=1}^N{\bm R}_i$, and $\overline{\bm S}=\frac{1}{N}\sum_{i=1}^{N}\bm S_i$. To confirm this, note that 
$\sum_{i=1}^N(\bm R_i-\bm S_i)^2$ reduces to $N(\bm R-\overline{\bm S})^2$ plus 
terms depending on the relative coordinates of 
$\bm R_i$'s. That is, letting each $\bm R_i$ shift to $\bm R_i+\bm v$, 
one only has to 
confirm that the difference, $\sum_{i=1}^N(\bm R_i-\bm S_i)^2-N(\bm R-\overline{\bm S})^2$, remains unchanged with respect to arbitrary $\bm v$: In fact 
$\sum_{i=1}^N (\bm R_i+\bm v-\bm S_i)^2-N(\bm R +\bm v-\overline{\bm S})^2
=\sum_{i=1}^N(\bm R_i-\bm S_i)^2-N(\bm R-\overline{\bm S})^2$. After all  
$\phi(\{\bm S\})$ factorizes into an 
intrinsic or translation-invariant wave function 
$\phi^{\rm in}(\{\bm S\})$ and the total c.m. wave function:
\begin{align}
\phi(\{\bm S\})=\phi^{\rm in}(\{\bm S\})\phi^{4N\beta}_{\overline{\bm S}}(\bm R),
\label{intri.cm.sepa}
\end{align}
which facilitates the calculation 
of c.m. motion-free (or intrinsic) matrix elements. For example, using the 
relation 
\begin{align}
\langle \phi(\{\bm S\})|\phi(\{\bm S'\})\rangle=\langle \phi^{\rm in}(\{\bm S\})|\phi^{\rm in}(\{\bm S'\})\rangle \langle \phi^{4N\beta}_{\overline{\bm S}}|\phi^{4N\beta}_{\overline{\bm S'}}\rangle,
\label{rel.full.intr}
\end{align}
where $\overline{\bm S'}=\frac{1}{N}\sum_{i=1}^{N}\bm S_i'$ and 
$\langle \phi^{4N\beta}_{\overline{\bm S}}|\phi^{4N\beta}_{\overline{\bm S'}}\rangle=e^{-N\beta (\overline{\bm S}- \overline{\bm S'})^2}
=e^{-\frac{\beta}{N}[\sum_{i=1}^N(\bm S_i -\bm S'_i)]^2}$,
and constraining $\overline{\bm S}=\overline{\bm S'}=\bm 0$,
I get the intrinsic matrix element  
$\langle \phi^{\rm in}(\{\bm S\})|\phi^{\rm in}(\{\bm S'\})\rangle$ by 
\begin{align}
\langle \phi^{\rm in}(\{\bm S\})|\phi^{\rm in}(\{\bm S'\})\rangle=\langle \phi(\{\bm S\})|\phi(\{\bm S'\})\rangle_{\overline{\bm S}=\overline{\bm S'}=\bm 0}.
\end{align}
The constraint, $\overline{\bm S}=\overline{\bm S'}=\bm 0$, is more 
conveniently realized by a translation of the system. That is, given a 
pair of sets 
$\{\bm S_i\}$ and $\{\bm S'_i\}$, I redefine a new pair of sets by shifting, 
$\{\bm S_i - \overline{\bm S}\}$ 
and $\{\bm S'_i - \overline{\bm S'}\}$, respectively. After this redefinition, 
no restriction on $\{\bm S_i\}$ and $\{\bm S'_i\}$ is necessary, but the constraint is automatically met. 

Table~\ref{spmelem} lists basic single-particle matrix elements of the GWPs.

\begin{table*}[h]
\caption{  Matrix element $\langle \phi^{\beta}_{\bm s}|{\cal O}|\phi^{\beta}_{\bm s'}\rangle$ of a single-particle operator ${\cal O}$ between the 
GWPs~(\ref{gwp}).  The matrix element is expressed as $M_{\cal O} e^{-\beta {\bm \sigma}_-^2}$.   Here 
$\bm \sigma_{\pm}=\frac{1}{2}(\bm s \pm \bm s')$  
and $\bm v$ is an arbitrary constant vector.}
\begin{tabular}{lcccl}
\hline\hline
 ${\cal O}$ &&&& $M_{\cal O}$ \\
\hline\hline
1 &&&& 1 \\
$\bm r$ &&&& $\bm \sigma_+$ \\
${\bm r}^2$ &&&& $\frac{3}{2\beta}+\bm \sigma_+^2$\\
${\cal Y}_{2\mu}(\bm r)$ &&&& ${\cal Y}_{2\mu}(\bm \sigma_+)$\\
${\cal Y}_{1\mu}(\bm r)$ &&&& ${\cal Y}_{1\mu}(\bm \sigma_+)$\\
${\bm r}^2{\cal Y}_{1\mu}(\bm r)$ &&&& $\big[\frac{5}{2\beta}+\bm \sigma_+^2\big]{\cal Y}_{1\mu}(\bm \sigma_+)$\\ 
$\bm v \cdot \bm r {\cal Y}_{1\mu}(\bm r)$ &&&& $\frac{1}{2 \beta}{\cal Y}_{1\mu}(\bm v)+\bm v\cdot \bm \sigma_+{\cal Y}_{1\mu}(\bm \sigma_+)$\\
${\cal Y}_{3\mu}(\bm r)$ &&&& ${\cal Y}_{3\mu}(\bm \sigma_+)$\\
${\bm p}$  &&&& $i\hbar \beta {\bm \sigma}_-$\\
${\bm p}^2$ &&&& $\hbar^2 \beta 
\big[\frac{3}{2}-\beta \bm \sigma_-^2\big]$\\
${\bm r}\cdot {\bm p}$ &&&& $i\hbar \big[\frac{3}{2}+\beta {\bm \sigma_+}\cdot {\bm \sigma_-}\big]$\\
${\cal Y}_{1\mu}(\bm r){\bm v}\cdot {\bm p}$  &&&& $i\hbar \big[\frac{1}{2}{\cal Y}_{1\mu}(\bm v)+\beta  \bm v\cdot \bm \sigma_- {\cal Y}_{1\mu}({\bm \sigma_+})\big]$   \\
${\cal Y}_{1\mu}(\bm r)\bm r\cdot \bm p$ &&&& $i\hbar \big[2{\cal Y}_{1\mu}({\bm \sigma}_+)+\frac{1}{2}{\cal Y}_{1\mu}({\bm \sigma}_-)+\beta \bm \sigma_+\cdot \bm \sigma_- {\cal Y}_{1\mu}({\bm \sigma}_+) \big]$ \\
%${\bm l}$ &&&& $i\hbar \beta ({\bm \sigma}_+\times {\bm \sigma}_-)$ \\
%${\cal Y}_{1\mu}(\bm r)\bm v\cdot{\bm l}$  &&&& $i\hbar \beta \big[\frac{1}{2\beta}{\cal Y}_{1\mu}(\bm v \times (\bm \sigma_+ - \bm \sigma_-))+\bm v\cdot (\bm \sigma_+\times \bm \sigma_-){\cal Y}_{1\mu}(\bm \sigma_+)\big]$ \\
%$\bm v\cdot{\bm l}{\cal Y}_{1\mu}(\bm r)$  &&&& $-i\hbar \beta \big[\frac{1}{2\beta}{\cal Y}_{1\mu}(\bm v \times (\bm \sigma_+ - \bm \sigma_-))+\bm v\cdot (\bm \sigma_+\times \bm \sigma_-){\cal Y}_{1\mu}(\bm \sigma_+)\big]$ \\
%${\bm l}^2$ &&&& $\hbar^2 \beta^2 \big[\frac{1}{\beta}(\bm \sigma_+^2 - \bm \sigma_-^2)+(\bm \sigma_+\cdot \bm \sigma_-)^2-\bm \sigma_+^2 \bm \sigma_-^2\big]$  \\
%${\bm l}{\cal Y}_{1\mu}(\bm r)\cdot {\bm l}$  &&&& $\hbar^2 \beta^2 \big[\frac{1}{\beta}\big\{(\bm \sigma_+^2-2\bm \sigma_-^2){\cal Y}_{1\mu}(\bm \sigma_+)+\bm \sigma_+\cdot \bm \sigma_- {\cal Y}_{1\mu}(\bm \sigma_-) \big\}+\big\{(\bm \sigma_+\cdot \bm \sigma_-)^2-\bm \sigma_+^2 \bm \sigma_-^2\big\}{\cal Y}_{1\mu}(\bm \sigma_+)\big]$ \\
\hline\hline
\end{tabular}
\label{spmelem}
\end{table*}

\subsection{Matrix elements between Slater determinants}
\label{cal.mat.slater.det}

\begin{center}
{\it I: Overlap}
\end{center}

The overlap of two Slater determinants is 
\begin{align}
\langle \phi(\{\bm S\})|\phi(\{\bm S'\})\rangle=
{\rm det}B,
\end{align}
where $B$ is a $4N\times 4N$ matrix and block-diagonal~\cite{brink66}:
\begin{align}
B=\left(\begin{array}{cccc}
b & 0 & 0 & 0 \\
0 & b & 0 & 0 \\
0 & 0 & b & 0 \\
0 & 0 & 0 & b \\
\end{array}
\right),\ \ \ \ \ 
B^{-1}=\left(\begin{array}{cccc}
b^{-1} & 0 & 0 & 0 \\
0 & b^{-1} & 0 & 0 \\
0 & 0 & b^{-1} & 0 \\
0 & 0 & 0 & b^{-1} \\
\end{array}
\right).
\label{overlapmatrix}
\end{align}
Here, $b$  is an $N \times N $ single-particle overlap matrix with 
its element $b_{ij}=\langle \phi_{\bm S_i}^{\beta}| \phi_{\bm S'_j}^{\beta} \rangle$, and $0$ is an $N \times N$ zero matrix. By letting $p$ denote a permutation of $1,2,\ldots, N$, $p=\left(\begin{array}{cccc}
1 & 2 & \cdots & N \\
p_1 & p_2 & \cdots & p_N\\
\end{array} \right)$, the determinant of $b$ is
\begin{align}
{\rm det}\, b=\sum_p\epsilon(p) b_{1p_1}\cdots b_{Np_N}=
\sum_p\epsilon(p)e^{-\frac{\beta}{4} \sum_{i=1}^N (\bm S_i-\bm S'_{p_i})^2},
\label{def.detb}
\end{align}
where $\sum_p$ extends over $N!$ permutations. 
The overlap of  two Slater determinants is 
\begin{align}
\langle \phi(\{\bm S\})|\phi(\{\bm S'\})\rangle=
\left(\sum_p\epsilon(p)e^{-\frac{\beta}{4} \sum_{i=1}^N (\bm S_i-\bm S'_{p_i})^2}\right)^4.
\label{ovlap.w.cm}
\end{align}
For $N=4$, ${\rm det}\, b$ contains $4!=24$ terms, and 
${\rm det}B=({\rm det}b)^4$ at most 17550 terms~\cite{count}, 
which are by far 
smaller than 16!=20922789888000 terms. To obtain an analytic form of the matrix element appears practicable for $N=4$.  

Owing to Eq.~(\ref{rel.full.intr}), 
I get the c.m. motion-free overlap matrix element by 
\begin{align}
\langle\phi^{\rm in}(\{\bm S\})|\phi^{\rm in}(\{\bm S'\})\rangle
=\left(\sum_p\epsilon(p)
e^{-\frac{\beta}{4} [\sum_{i=1}^N (\bm S_i-\bm S'_{p_i})^2-N(\overline{\bm S}- \overline{\bm S'})^2] }\right)^4.
\label{ovlap.wo.cm}
\end{align}
Comparing Eqs.~(\ref{ovlap.w.cm}) and~(\ref{ovlap.wo.cm}) again confirms that 
imposing $\overline{\bm S}=\overline{\bm S'}=\bm 0$ amounts to 
renaming the sets, $\{\bm S_i\}$ and $\{\bm S_i'\}$, to $\{\bm S_i - \overline{\bm S}\}$ 
and $\{\bm S'_i - \overline{\bm S'}\}$, respectively. Note that under this constraint the following equation holds,   
\begin{align}
\sum_{i=1}^N (\bm S_i\pm \bm S'_{p_i})^2 \to \sum_{i=1}^N(\bm S_i-\overline{\bm S}\pm (\bm S'_{p_i}-\overline{\bm S'}))^2
=\sum_{i=1}^N(\bm S_i\pm \bm S'_{p_i})^2-N(\overline{\bm S}\pm \overline{\bm S'})^2.
\end{align}

I give a general prescription to set the constraint for the matrix element 
of such a 
one-body operator, ${\cal O}$, that has no dependence on the 
spin-isospin coordinates. If $\langle \phi_{\bm S_i}^{\beta}|{\cal O}|\phi_{\bm S'_j}^{\beta}\rangle$ is given by $O(\bm S_i, \bm S_j')b_{ij}$, 
\begin{align}
\big\langle \phi(\{\bm S \})|\sum_{i=1}^{4N}{\cal O}_i|\phi(\{\bm S'\})\big\rangle
&=4({\rm det} b)^3 \sum_{i=1}^N \sum_p \epsilon(p)b_{1p_1}\cdots O(\bm S_i, \bm S'_{p_i})b_{ip_i}\cdots b_{Np_N}\notag \\
&=\left(\sum_p\epsilon(p)e^{-\frac{\beta}{4} \sum_{i=1}^N (\bm S_i-\bm S'_{p_i})^2}\right)^3  \sum_p \epsilon(p)  e^{-\frac{\beta}{4} \sum_{i=1}^N (\bm S_i-\bm S'_{p_i})^2}O_p(\bm S, \bm S'),
\end{align}
where
\begin{align}
O_p(\bm S, \bm S')=4\sum_{i=1}^N O(\bm S_i, \bm S'_{p_i}).
\label{single.op.perm}
\end{align}
$O_p(\bm S, \bm S')$ is obtained by using $O(\bm S_i, \bm S_j')$ available in 
Table~\ref{spmelem}. The matrix element 
$\big\langle \phi(\{\bm S \})|\sum_{i=1}^{4N}{\cal O}_i|\phi(\{\bm S'\})\big\rangle_{\overline{\bm S}=\overline{\bm S'}=\bm 0}$ is obtained by replacing 
$\{\bm S_i\}$ and $\{\bm S_i'\}$ by $\{\bm S_i - \overline{\bm S}\}$ 
and $\{\bm S'_i - \overline{\bm S'}\}$, respectively:
\begin{align}
\big\langle \phi(\{\bm S \})|\sum_{i=1}^{4N}{\cal O}_i|\phi(\{\bm S'\})\big\rangle_{\overline{\bm S}=\overline{\bm S'}=\bm 0}&=
\left(\sum_p\epsilon(p)
e^{-\frac{\beta}{4} [\sum_{i=1}^N (\bm S_i-\bm S'_{p_i})^2-N(\overline{\bm S}- \overline{\bm S'})^2] }\right)^3 \notag \\
&\times \sum_p \epsilon(p)
e^{-\frac{\beta}{4} [\sum_{i=1}^N (\bm S_i-\bm S'_{p_i})^2-N(\overline{\bm S}- \overline{\bm S'})^2] }O_p(\bm S, \bm S')_{\overline{\bm S}=\overline{\bm S'}=\bm 0}
\end{align}
with
\begin{align}
O_p(\bm S, \bm S')_{\overline{\bm S}=\overline{\bm S'}=\bm 0}=4\sum_{i=1}^N O(\bm S_i-\overline{\bm S}, \bm S'_{p_i}-\overline{\bm S'}).
\end{align}
Further consideration is  needed to eliminate a c.m. 
contribution that the operator ${\cal O}$ may produce, as shown below.\\

\begin{center}
{\it II: Kinetic energy (KE) and squared radius (SR)}
\end{center}

For ${\cal O}_i=\frac{{\bm p_i}^2}{2m_N}$ (KE) or ${\bm r_i}^2$ (SR), $O_p(\bm S, \bm S')$ of Eq.~(\ref{single.op.perm}) reads 
\begin{align}
KE_p(\bm S, \bm S')=\frac{\hbar^2 \beta}{m_N}
\big[3N-\frac{\beta}{2}\sum_{i=1}^N(\bm S_i-\bm S_{p_i}')^2\big],\ \ \ \ \ 
SR_p(\bm S, \bm S')=\frac{6N}{\beta}+\sum_{i=1}^N(\bm S_i+\bm S_{p_i}')^2.
\end{align}
Here, $m_N$ is the nucleon mass, an average of the neutron mass and the 
proton mass. 
Both the kinetic energy and the squared radius are separated into 
a sum of operators depending on the intrinsic coordinates and the c.m. 
coordinate:  
\begin{align}
\sum_{i=1}^{4N} \frac{1}{2m_N}  \big(\bm p_i-\frac{1}{4N}\bm P \big)^2=\sum_{i=1}^{4N} \frac{ \bm p_i^2}{2m_N}- \frac{\bm P^2}{2(4N)m_N},\ \ \ \ \  
\sum_{i=1}^{4N}(\bm r_i-\bm R)^2=\sum_{i=1}^{4N} \bm r_i^2-4N \bm R^2,
\end{align}
where $\bm P=\sum_{i=1}^{4N}\bm p_i$ is the total momentum. The matrix elements of $\bm P^2$ and $\bm R^2$ with the c.m. wave function constrained by 
$\overline{\bm S}=\overline{\bm S'}=\bm 0$ are 
obtained from Table~\ref{spmelem}, leading to the desired matrix elements as follows:
\begin{align}
& \big\langle\phi^{\rm in}(\{\bm S\})|\sum_{i=1}^{4N} \frac{1}{2m_N}  \big(\bm p_i-\frac{1}{4N}\bm P \big)^2|\phi^{\rm in}(\{\bm S'\})\big\rangle
 =\big\langle \phi(\{\bm S\})|\sum_{i=1}^{4N}\frac{\bm p_i^2}{2m_N}|\phi(\{\bm S'\})\big\rangle_{\overline{\bm S}=\overline{\bm S'}=\bm 0}
-\frac{3\hbar^2 \beta}{4m_N}
\langle\phi^{\rm in}(\{\bm S\})|\phi^{\rm in}(\{\bm S'\})\rangle , \\
&\big\langle\phi^{\rm in}(\{\bm S\})|\sum_{i=1}^{4N}({\bm r}_i-{\bm R})^2|\phi^{\rm in}(\{\bm S'\})\big\rangle
 =\big\langle\phi(\{\bm S\})|\sum_{i=1}^{4N}{\bm r_i}^2|\phi(\{\bm S'\})\big\rangle_{\overline{\bm S}=\overline{\bm S'}=\bm 0} -\frac{3}{2 \beta}
\langle\phi^{\rm in}(\{\bm S\})|\phi^{\rm in}(\{\bm S'\})\rangle.
\end{align}

\begin{center}
{\it III: Electric quadrupole (QD)}
\end{center}

The $E2$ operator is given by $\sum_{i=1}^{4N}(\frac{1}{2}-t_3(i)){\cal Y}_{2\mu}(\bm r_i)$, where $t_3$ is the $z$ component 
of the nucleon isospin. 
Since the $t_3$-dependent term has no contribution in the $N \alpha$-particle model, 
one only has to consider the IS part of the $E2$ operator, 
$\frac{1}{2}\sum_{i=1}^{4N}{\cal Y}_{2\mu}(\bm r_i)$. The use of Table~\ref{spmelem} gives 
\begin{align}
QD_p(\bm S, \bm S')
=\frac{1}{2}\sum_{i=1}^N{\cal Y}_{2\mu}(\bm S_i+\bm S'_{p_i}).
\end{align}
The $E2$ operator factorizes into the intrinsic and c.m. parts (see Eq.~(6.108)~\cite{book98}),
\begin{align}
\sum_{i=1}^{4N}{\cal Y}_{2\mu}(\bm r_i-\bm R)=\sum_{i=1}^{4N}{\cal Y}_{2\mu}(\bm r_i)-4N{\cal Y}_{2\mu}(\bm R).
\end{align}
${\cal Y}_{2\mu}(\bm R)$ has no contribution to $\langle \phi^{4N\beta}_{\overline{\bm S}}(\bm R)| {\cal Y}_{2\mu}(\bm R)|\phi^{4N\beta}_{\overline{\bm S'}}(\bm R)\rangle_{\overline{\bm S}=\overline{\bm S'}=\bm 0}=0$, and the $E2$ matrix element reads as 
\begin{align}
\langle\phi^{\rm in}(\{\bm S\})|\frac{1}{2}\sum_{i=1}^{4N}{\cal Y}_{2\mu}({\bm r}_i-{\bm R})|\phi^{\rm in}(\{\bm S'\})\rangle= \langle \phi(\{\bm S\})|\frac{1}{2}\sum_{i=1}^{4N}{\cal Y}_{2\mu}(\bm r_i) |\phi(\{\bm S'\})\rangle_{\overline{\bm S}=\overline{\bm S'}=\bm 0}.
\end{align}

\begin{center}
{\it IV: Isoscalar electric dipole (DP)}
\end{center}

The matrix element $A_{\rm IS}$ of Eq.~(\ref{ISE1mat}) is calculated here. 
The IS $E1$ operator $E1({\rm IS})$ comprises three terms up to 
the first order beyond the long-wavelength approximation.  The spin-dependent term among them has no contribution because $\phi(\{\bm S \})$ has 
zero spin.  There are two choices to specify the remaining terms~\cite{baye12}: A combination of either $-e\frac{k^2}{10}{\bm r}^2{\cal Y}_{1\mu}(\bm r)$ and $\frac{e\hbar k}{8m_pc}\frac{2i}{\hbar}{\cal Y}_{1\mu}({\bm r})\bm r\cdot {\bm p}$ or 
$-e\frac{k^2}{60}{\bm r}^2{\cal Y}_{1\mu}(\bm r)$ and $\frac{e k}{8m_pc \hbar}\frac{2}{3}{\bm l}{\cal Y}_{1\mu}({\bm r})\cdot {\bm l}$, where $m_p$ is the proton mass and 
$\frac{1}{m_p}$ can be replaced by $\frac{1}{m_N}$ because 
$\frac{1}{m_p}\approx \frac{1}{m_N}(1+6.89\times 10^{-4})$ and $\bm l=\bm r \times \bm p$.  The first choice 
is adopted because of its simplicity. 
\\

\noindent
\underline{${\bm r}^2{\cal Y}_{1\mu}(\bm r)$ term,  {\it DP1}}
\begin{align}
&DP{\it 1}_p(\bm S, \bm S')=4\sum_{i=1}^N \Big[\frac{5}{4\beta}+\frac{1}{8}(\bm S_i+\bm S_{p_i}')^2\Big]{\cal Y}_{1\mu}(\bm S_i+\bm S_{p_i}')=
\frac{5N}{\beta}{\cal Y}_{1\mu}(\overline{\bm S}+\overline{\bm S'})+
\frac{1}{2}\sum_{i=1}^N(\bm S_i+\bm S_{p_i}')^2{\cal Y}_{1\mu}(\bm S_i+\bm S_{p_i}').
\end{align}
${\cal Y}_{1\mu}(\overline{\bm S}+\overline{\bm S'})$ has no contribution 
to $DP{\it 1}_p(\bm S, \bm S')_{\overline{\bm S}=\overline{\bm S'}=\bm 0}$. Using the identity
\begin{align}
&\sum_{i=1}^{4N}{{\bm r}_i}^2{\cal Y}_{1\mu}(\bm r_i)=\sum_{i=1}^{4N}\Big[(\bm r_i-\bm R)^2 {\cal Y}_{1\mu}(\bm r_i -\bm R)
+2\bm R\cdot (\bm r_i-\bm R){\cal Y}_{1\mu}(\bm r_i-\bm R)
+ (\bm r_i-\bm R)^2{\cal Y}_{1\mu}(\bm R)+{\bm R}^2{\cal Y}_{1\mu}(\bm R)\Big],
\label{rrY.1} 
\end{align}
and noting that all the c.m. matrix elements vanish as in the case of the $E2$ operator,  I get the desired matrix element 
\begin{align}
\big\langle \phi^{\rm in}(\{\bm S\})|\sum_{i=1}^{4N}(\bm r_i-\bm R)^2{\cal Y}_{1\mu}(\bm r_i -\bm R)|\phi^{\rm in}(\{\bm S'\})\big\rangle
=\big\langle \phi(\{\bm S\})|\sum_{i=1}^{4N} {\bm r}_i^2{\cal Y}_{1\mu}(\bm r_i) |\phi(\{\bm S'\})\big\rangle_{\overline{\bm S}=\overline{\bm S'}=\bm 0}. 
\end{align}
\\

\noindent
\underline{${\cal Y}_{1\mu}({\bm r})\bm r\cdot {\bm p}$ term,  {\it DP2}}
\begin{align}
&DP{\it 2}_p(\bm S, \bm S')=4i\hbar \sum_{i=1}^N\Big[{\cal Y}_{1\mu}(\bm S_i+\bm S_{p_i}')+
\frac{1}{4}{\cal Y}_{1\mu}(\bm S_i-\bm S_{p_i}')+\frac{\beta}{8}(\bm S_i^2-\bm S_{p_i}'^2){\cal Y}_{1\mu}(\bm S_i+\bm S_{p_i}')\Big].
\end{align}
Neither $\sum_{i=1}^N{\cal Y}_{1\mu}(\bm S_i+\bm S_{p_i}')$ nor $\sum_{i=1}^N{\cal Y}_{1\mu}(\bm S_i-\bm S_{p_i}')$ contributes 
to $DP{\it 2}_p(\bm S, \bm S')_{\overline{\bm S}=\overline{\bm S'}=\bm 0}$. Using the decomposition 
\begin{align}
\sum_{i=1}^{4N}{\cal Y}_{1\mu}({\bm r}_i){\bm r}_i\cdot{\bm p}_i&=\sum_{i=1}^{4N}\Big[{\cal Y}_{1\mu}({\bm r}_i-{\bm R})({\bm r}_i-{\bm R})\cdot({\bm p}_i-\frac{1}{4N}{\bm P})+\frac{1}{4N}{\cal Y}_{1\mu}({\bm r}_i-{\bm R})({\bm r}_i-{\bm R})\cdot {\bm P}\notag \\
&+{\cal Y}_{1\mu}({\bm r}_i-{\bm R}){\bm R}\cdot ({\bm p}_i-\frac{1}{4N}{\bm P})+{\cal Y}_{1\mu}(\bm R)({\bm r}_i-{\bm R})\cdot ({\bm p}_i-\frac{1}{4N}{\bm P})+
\frac{1}{4N}{\cal Y}_{1\mu}(\bm R){\bm R}\cdot {\bm P}\Big],
\end{align}
and noting again that all the c.m. matrix elements vanish leads to 
the desired  matrix element: 
\begin{align}
\big\langle \phi^{\rm in}(\{\bm S\})|\sum_{i=1}^{4N}{\cal Y}_{1\mu}({\bm r}_i-{\bm R})({\bm r}_i-{\bm R})\cdot({\bm p}_i-\frac{1}{4N}{\bm P}) |\phi^{\rm in}(\{\bm S'\})\big\rangle
=\big\langle \phi(\{\bm S\})|\sum_{i=1}^{4N} {\cal Y}_{1\mu}(\bm r_i){\bm r}_i\cdot{\bm p}_i |\phi(\{\bm S'\})\big\rangle_{\overline{\bm S}=\overline{\bm S'}=\bm 0}. 
\end{align}

The matrix elements of both DP1 and DP2 operators consist of terms with 
$\bm S^2 {\cal Y}_{1\mu}(\bm S)$ type. To give a rough estimate of their 
relative importance I compare the coefficients 
appearing in $A_{\rm IS}$. The coefficients divided by $\beta$ are 
$-\frac{k^2}{20\beta}$ and $-\frac{\hbar k}{8m_Nc}$. For $k=0.036$\, fm$^{-1}$, 
$\beta=0.54$\, fm$^{-2}$, they are respectively $-1.2\times 10^{-4}$ and 
$-9.5\times 10^{-4}$. 

\begin{center}
{\it V: Electric octupole (OP)}
\end{center}

The $E3$ operator is  $\sum_{i=1}^{4N}OP_i$ with $OP_i=(\frac{1}{2}-t_3(i)){\cal Y}_{3\mu}({\bm r}_i)$. As in the case of  the $E2$ operator, only its IS part contributes, leading to 

\begin{align}
\big\langle \phi(\{\bm S\})|\sum_{i=1}^{4N}\frac{1}{2}{\cal Y}_{3\mu}({\bm r}_i) |\phi(\{\bm S'\})\big\rangle
=\left( \sum_p \epsilon(p)e^{-\frac{\beta}{4}\sum_{i=1}^N(\bm S_i-\bm S'_{p_i})^2}\right)^3 \sum_p \epsilon(p)e^{-\frac{\beta}{4}\sum_{i=1}^N(\bm S_i-\bm S'_{p_i})^2} OP_p(\bm S, \bm S'),
\end{align}
where
\begin{align}
OP_p(\bm S, \bm S')=\frac{1}{2}4\sum_{i=1}^N {\cal Y}_{3\mu}\big(\frac{1}{2}(\bm S_i+\bm S'_{p_i})\big)=\frac{1}{4}
\sum_{i=1}^N {\cal Y}_{3\mu}(\bm S_i+\bm S'_{p_i}).
\end{align} 
Using the decomposition (see Eq.~(6.105) of  Ref.~\cite{book98})
\begin{align}
\sum_{i=1}^{4N}{\cal Y}_{3\mu}(\bm r_i)=\sum_{i=1}^{4N}{\cal Y}_{3\mu}(\bm r_i-\bm R + \bm R)
=\sum_{l=0}^3 \sqrt{\frac{4\pi \  7!}{(2l+1)!(7-2l)!}}\Big[\sum_{i=1}^{4N}{\cal Y}_l(\bm r_i-\bm R) \times {\cal Y}_{3-l}(\bm R)\Big]_{3\mu}
\end{align}
and noting that ${\cal Y}_{3-l}(\bm R)$ has a non-vanishing matrix element only for $l=3$ under the constraint  $\overline{\bm S}=
\overline{\bm S'}=\bm 0$, I obtain the desired $E3$  matrix elemnt as follows:
\begin{align}
\big\langle \phi^{\rm in}(\{\bm S\})|\sum_{i=1}^{4N}\frac{1}{2}{\cal Y}_{3\mu}({\bm r}_i-\bm R) |\phi^{\rm in}(\{\bm S'\})\big\rangle
=\big\langle \phi(\{\bm S\})|\sum_{i=1}^{4N}\frac{1}{2}{\cal Y}_{3\mu}({\bm r}_i) |\phi(\{\bm S'\})\big\rangle_{\overline{\bm S}=\overline{\bm S'}=\bm 0}.
\end{align}

\begin{center}
{\it VI: Two-body potential}
\end{center}

Finally I take up a two-body central potential, $\sum_{i<j=1}^{4N}V_{ij}$, where 
$V_{ij} = \delta(\bm r_i-\bm r_j-\bm r)O_{ij}$ and $O_{ij}$ is specific to 
the potential. E.g., ${O}_{ij}=w+bP^{\sigma}_{ij}-hP^{\tau}_{ij}-mP^{\sigma}_{ij}P^{\tau}_{ij}$ for the nuclear potential and $O_{ij}=\delta_{i,p}\delta_{j,p}$ for 
the Coulomb potential. Here $P^{\sigma}_{ij}$ and $P^{\tau}_{ij}$ are the spin- and isospin-exchange operators, respectively. Since each $\phi^{\beta}_{{\bm S}_i}$ is occupied by 
four nucleons, the two-body matrix element reads as (see Ref.~\cite{brink66})
\begin{align}
&\big\langle \phi(\{\bm S\})|\sum_{i<j=1}^{4N}V_{ij}|\phi(\{\bm S'\})\big\rangle\notag \\ 
&\quad =({\rm det}\,b)^4 \sum_{i,j=1}^N\sum_{k,l=1}^N
\langle \phi^{\beta}_{{\bm S}_i}(\bm r_1)\phi^{\beta}_{{\bm S}_j}(\bm r_2)|\delta(\bm r_1-\bm r_2-\bm r) |\phi^{\beta}_{\bm S'_k}(\bm r_1)\phi^{\beta}_{\bm S'_l}(\bm r_2)\rangle \big(X_d b^{-1}_{\ ki}b^{-1}_{\ lj}+X_eb^{-1}_{\ kj}b^{-1}_{\ li}\big)\notag \\
&\quad =   ({\rm det}\,b)^4 \Big(\frac{\beta}{2\pi}\Big)^{\frac{3}{2}} \sum_{i,j=1}^N\sum_{k,l=1}^N 
e^{-\frac{\beta}{2}(\bm r-\frac{1}{2}(\bm S_i-\bm S_j+\bm S'_k-\bm S'_l))^2}b_{ik}b_{jl}
\big(X_d b^{-1}_{\ ki}b^{-1}_{\ lj}+X_eb^{-1}_{\ kj}b^{-1}_{\ li}\big),
\label{two-body.cf}
\end{align}
where $b^{-1}_{\ ki}$ is the $(k,i)$ element of the inverse matrix of $b$, and $X_d$ and $X_e$ are 
\begin{align}
&{\rm for\ the\ nuclear\ potential}\ \ \ \ \ \ \ \, X_d=8w+4b-4h-2m,\ \ \ \ \ X_e=8m+4h-4b-2w,\notag \\
&{\rm for\ the\ Coulomb\ potential}\ \ \ \ \ X_d=4,\ \ \ \ \ X_e=-2.
\end{align}
It is convenient to express $b^{-1}_{\  ki}$ as 
\begin{align}
&b^{-1}_{\  ki}=\frac{1}{{\rm det}\, b}\frac{1}{b_{ik}}\sum_{p}\epsilon(p)b_{1p_1}\cdots b_{Np_N} \delta_{p_i,k},
\end{align}
where $p$ is constrained to $p_i=k$. By interchanging $i, j$ for the $X_e$ term, 
the matrix element~(\ref{two-body.cf}) is recast to 
\begin{align}
&\big\langle \phi(\{\bm S\})|\sum_{i<j=1}^{4N}V_{ij}|\phi(\{\bm S'\})\big\rangle
\notag \\
&=({\rm det}\,b)^2  \sum_{i,j=1}^N \sum_{k,l=1}^N
\left(\sum_p \epsilon(p) e^{-\frac{\beta}{4}\sum_{m=1}^N (\bm S_m-\bm S'_{p_m})^2} \delta_{p_i,k}\right) \left(\sum_{q} \epsilon(q) 
e^{-\frac{\beta}{4}\sum_{n=1}^N (\bm S_n-\bm S'_{q_n})^2} \delta_{q_j,l}\right) \notag \\
&\times  \Big(\frac{\beta}{2\pi}\Big)^{\frac{3}{2}}  
\left(X_d\, e^{-\frac{\beta}{2}(\bm r-\frac{1}{2}(\bm S_i-\bm S_j+\bm S'_k-\bm S'_l))^2}
+X_e \, e^{-\frac{\beta}{2}(\bm r+\frac{1}{2}(\bm S_i-\bm S_j-\bm S'_k+\bm S'_l))^2-\frac{\beta}{2}(\bm S_i-\bm S_j)\cdot (\bm S'_k-\bm S'_l)}\right).
\label{me.pot.SD.cm.free}
\end{align}
Note that $e^{-\frac{\beta}{4}\sum_{m=1}^N (\bm S_m-\bm S'_{p_m})^2} \delta_{p_i,k}$ reads 
$e^{-\frac{\beta}{4}\sum_{m\neq i}^N (\bm S_m-\bm S'_{p_m})^2-\frac{\beta}{4}(\bm S_i -\bm S'_k)^2} \delta_{p_i,k}$.  Replacing the sum of type, 
$\sum_{k=1}^N (\bm S_k-\bm S'_{p_k})^2$, 
by $\sum_{k=1}^N (\bm S_k-\bm S'_{p_k})^2-N(\overline{\bm S}-\overline{\bm S'})^2$ as before, I get an intrinsic matrix element, $\big\langle \phi^{\rm in}(\{\bm S\})|\sum_{i<j=1}^{4N}V_{ij}|\phi^{\rm in}(\{\bm S'\})\big\rangle$. 

In order to obtain the matrix element for the Gaussian potential, $e^{-\rho(\bm r_i-\bm r_j)^2}O_{ij}$, one only has to multiply Eq.~(\ref{me.pot.SD.cm.free}) by 
$e^{-\rho {\bm r}^2}$ and to integrate over $\bm r$,   
which results in replacing the last line of Eq.~(\ref{me.pot.SD.cm.free}) by 
\begin{align}
\Big(\frac{\beta}{\beta+2\rho}\Big)^{\frac{3}{2}} \left(X_d \, e^{-\frac{\beta \rho}{4(\beta+2\rho)}(\bm S_i-\bm S_j+\bm S'_k-\bm S'_l)^2}+X_e \, e^{-\frac{\beta \rho}{4(\beta+2\rho)}(\bm S_i-\bm S_j-\bm S'_k+\bm S'_l)^2 -\frac{\beta}{2}(\bm S_i-\bm S_j)\cdot (\bm S'_k-\bm S'_l)} \right).
\label{me.gauss}
\end{align} 
The matrix element for the Coulomb potential, 
$\frac{1}{|\bm r_i-\bm r_j|}O_{ij}$, is readily derived from the one of the 
Gaussian potential with a simple replacement. Using $\frac{1}{|\bm r_i-\bm r_j|}=\frac{2}{\sqrt{\pi}}
\int_0^{\infty}d\rho\, e^{-\rho^2(\bm r_i-\bm r_j)^2}$ and noting the $\rho$-dependence in Eq.~(\ref{me.gauss}) enables one to get the Coulomb matrix 
element through the following integration
\begin{align}
\frac{2}{\sqrt{\pi}}\int_0^{\infty}d\rho\, \Big(\frac{\beta}{\beta+2\rho^2}\Big)^{\frac{3}{2}}\,e^{-\frac{\beta \rho^2}{4(\beta+2\rho^2)}\bm v^2}
=\sqrt{\frac{2\beta}{\pi}}\int_0^1 dx\, e^{-\gamma^2 x^2},\ \ \ \ \ \gamma=\sqrt{\frac{\beta}{8}}v,
\end{align}
where $\bm v$ stands for $\bm S_i-\bm S_j+\bm S'_k-\bm S'_l$ for the $X_d$ 
term and $\bm S_i-\bm S_j-\bm S'_k+\bm S'_l$ for the $X_e$ term, respectively. 
The integral $\int_0^1 dx\, e^{-\gamma^2 x^2}$ can be well approximated  
by a few Gaussians, $\sum_i w_i e^{-x_i^2 \gamma^2}$.

\section{Isospin impurity of $\alpha$ particle and electric dipole matrix element}
\label{isf.alpha}

The $T=1$ impurity component of $\alpha$ particle, 
$\phi_{\alpha}^{(1)}$ in Eq.~(\ref{alpha*}), is assumed to 
have $L^{\pi}=0^+, \, S=0$. 
It is constructed by a combination of $^3$H($t$)+$p$ and 
$^3$He($h$)+$n$ two-cluster configurations as in Ref.~\cite{arai96}, where 
the isospin mixing of $\alpha$ particle is found to play a vital role in 
accounting for 
the beta decay of the $^9$Li ground-state to the $^9$Be ground-state.

\subsection{$T=1$ impurity of $\alpha$ particle}
\label{determination.epsilon}

The spin-isospin part of $t$ with $M_S=\frac{1}{2}$, the $z$ component of 
the spin, is given by 
\begin{align}
\Omega_{\frac{1}{2}}(t,123)=\frac{1}{\sqrt{3!}}\left|
\begin{array}{ccc}
\chi_1(1)&\chi_2(1)&\chi_3(1)\\
\chi_1(2)&\chi_2(2)&\chi_3(2)\\
\chi_1(3)&\chi_2(3)&\chi_3(3)\\
\end{array}\right|.
\label{t.spin-isospin}
\end{align}
$\Omega_{-\frac{1}{2}}(t,123)$ is defined by replacing $\chi_3$ by $\chi_4$. 
The $tp$ two-cluster configuration with $\pi=+$ and $S=0$ reads 
\begin{align}
\phi(tp,\bm q)
=\phi^{\rm orb}_{\bm q}(123,4)\frac{1}{\sqrt{2}}
\Big(\Omega_{\frac{1}{2}}(t,123)\chi_4(4)-
\Omega_{-\frac{1}{2}}(t,123)\chi_3(4)\Big),
\label{tp}
\end{align}
where the orbital part, $\phi^{\rm orb}_{\bm q}(123,4)$, is defined by the GWPs as 
\begin{align}
\phi^{\rm orb}_{\bm q}(123,4)=\frac{1}{2}\Big[\prod_{i=1}^3\phi^{\beta}_{\frac{1}{4}\bm q}(\bm r_i)\phi^{\beta}_{-\frac{3}{4}\bm q}(\bm r_4)+(\bm q \to -\bm q)\Big].
\end{align}
The $t$ cluster is centered at $\frac{1}{4}\bm q$, while  
$p$ is at $-\frac{3}{4}\bm q$. They are separated by 
$q=|\bm q|$. Adding the configuration 
with $\bm q \to -\bm q$ assures the positive parity of $\phi^{\rm orb}_{\bm q}(123,4)$. 
Similarly the spin-isospin part  of $h$ with $M_S=\frac{1}{2}$, 
$\Omega_{\frac{1}{2}}(h,123)$, is defined by replacing $\chi_1, \chi_2, \chi_3$ in Eq.~(\ref{t.spin-isospin}) by $\chi_3, \chi_4, \chi_1$, respectively. 
$\Omega_{-\frac{1}{2}}(h,123)$ is defined by replacing $\chi_1$ by $\chi_2$ in 
$\Omega_{\frac{1}{2}}(h,123)$. 
The $hn$ two-cluster configuration with $\pi=+$ and $S=0$ is
\begin{align}
\phi(hn,\bm q)
=\phi^{\rm orb}_{\bm q}(123,4)
\frac{1}{\sqrt{2}}\Big(\Omega_{\frac{1}{2}}(h,123)\chi_2(4)-
\Omega_{-\frac{1}{2}}(h,123)\chi_1(4)\Big).
\label{hn}
\end{align}
The isospin functions in Eqs.~(\ref{tp}) and~(\ref{hn}) contain both 
$T=0$ and 1 components. The $T=1$ state is given by
\begin{align}
\phi^{\rm orb}_{\bm q}(123,4)\Omega(123,4)\equiv \frac{1}{\sqrt{2}}[\phi(hn, \bm q)-\phi(tp,\bm q)]
\end{align}
with
\begin{align}
&\Omega(123,4)=\frac{1}{2}\Big(\Omega_{\frac{1}{2}}(h,123)\chi_2(4)-
\Omega_{-\frac{1}{2}}(h,123)\chi_1(4)-\Omega_{\frac{1}{2}}(t,123)\chi_4(4)+
\Omega_{-\frac{1}{2}}(t,123)\chi_3(4)\Big).
\label{def.Omega}
\end{align}

By its construction, $\phi^{\rm orb}_{\bm q}(123,4)\Omega(123,4)$ 
is antisymmetric 
with respect to the permutations of 1, 2, and 3, more precisely, antisymmetric 
in the spin-isospin space and symmetric in the orbital space.  
A fully antisymmetrized $T=1$ state, denoted $\phi^{(1)}_{\alpha}(\bm q)$, 
is constructed by acting ${\cal A}_4$ on $\phi^{\rm orb}_{\bm q}(123,4)\Omega(123,4)$: 
\begin{align}
&\phi^{(1)}_{\alpha}(\bm q)=\frac{6}{\sqrt{4!}}[1-(1,4)-(2,4)-(3,4)]
\phi^{\rm orb}_{\bm q}(123,4)\Omega(123,4),
\label{alpha.dist}
\end{align}
where, e.g.,  $(1,4)$ is the transposition of exchanging 1 and 4. To perform the transpositions, 
I rewrite $\Omega(123,4)$ to a representation, $|(S_{12}S_{34})SM_S, (T_{12}T_{34})TM_T\rangle$, where, e.g., $S_{12}$ ($T_{12}$) stands for the spin (isospin) resulting from coupling the spins (isospins) of the first nucleon and the 
second nucleon: 
\begin{align}
\Omega(123,4)=\frac{1}{\sqrt{3!}}\big\{|(00)00, (10)10\rangle - \sqrt{2}|(00)00,(11)10\rangle + \sqrt{3}|(11)00,(01)10\rangle \big\}.
\label{Omega123,4}
\end{align}
The action of $(1,4)$ is done by using (unitary) $R(3)$ 9$j$-coefficients~\cite{book98} that lead, e.g., in the spin part, to   
\begin{align}
(1,4)|(S_{12}S_{34})SM_S\rangle= (-1)^{S}\sum_{S_{12}' S_{34}'}\left[
\begin{array}{ccc}
\frac{1}{2}& \frac{1}{2}& S_{12}' \\
\frac{1}{2}& \frac{1}{2}& S_{34}' \\
S_{34}& S_{12} & S\\
\end{array}
\right]|(S_{12}'S_{34}')SM_S\rangle.
\label{racah}
\end{align}
$\Omega(123,4)$ is found to be subject to the following change under the transpositions:
\begin{align}
(1,4)\Omega(123,4)=\frac{1}{\sqrt{3!}}\big\{&|(00)00,(01)10\rangle-\sqrt{2}|(00)00,(11)10\rangle +\sqrt{3}|(11)00,(10)10\rangle  \big\},\notag \\
(2,4)\Omega(123,4)=\frac{1}{\sqrt{3!}}\big\{&-|(00)00,(01)10\rangle-\sqrt{2}|(00)00,(11)10\rangle -\sqrt{3}|(11)00,(10)10\rangle \big\},\notag \\
(3,4)\Omega(123,4)=\frac{1}{\sqrt{3!}}\big\{&|(00)00, (10)10\rangle +\sqrt{2}|(00)00,(11)10\rangle +\sqrt{3}|(11)00,(01)10\rangle \big\}.
\end{align}
Here, acting $(2,4)$ is most easily done by noting $(2,4)=(1,2)(1,4)(1,2)$.  
After all, I get
\begin{align}
\phi^{(1)}_{\alpha}(\bm q)&=\frac{1}{2} \phi^{\rm orb}_{\bm q}(423,1)
\big\{-|(00)00,(01)10\rangle+\sqrt{2}|(00)00,(11)10\rangle -\sqrt{3}|(11)00,(10)10\rangle
\big\}\notag \\
&+\frac{1}{2}\phi^{\rm orb}_{\bm q}(143,2)\big\{|(00)00,(01)10\rangle+\sqrt{2}|(00)00,(11)10\rangle +\sqrt{3}|(11)00,(10)10\rangle \big\}\notag \\
&+\frac{1}{2}\phi^{\rm orb}_{\bm q}(124,3)\big\{-|(00)00,(10)10\rangle-\sqrt{2}|(00)00,(11)10\rangle -\sqrt{3}|(11)00,(01)10\rangle  \big\}\notag \\
&+\frac{1}{2}\phi^{\rm orb}_{\bm q}(123,4) \big\{|(00)00,(10)10\rangle -\sqrt{2}|(00)00,(11)10\rangle +\sqrt{3}|(11)00,(01)10\rangle \big\}\notag \\
&\equiv \frac{1}{2}\sum_{i=1}^4\phi_{\bm q}(i)\Omega_i.
\label{impurity.state}
\end{align}
Here, e.g., $\phi_{\bm q}(1)=\phi_{\bm q}^{\rm orb}(423,1)$ is symmetric with respect to the exchange of $\bm r_j$'s other than $\bm r_1$, and $\Omega_1=-|(00)00,(01)10\rangle+\sqrt{2}|(00)00,(11)10\rangle -\sqrt{3}|(11)00,(10)10\rangle$. Not all of $\Omega_i$'s are independent, but $\sum_{i=1}^4 \Omega_i=0$.

$\phi^{(1)}_{\alpha}(\bm q)$ contains the c.m. motion, $\phi^{4\beta}_{\bm 0}(\bm R)$, with $\bm R=\frac{1}{4}\sum_{i=1}^4\bm r_i$. By removing it  
from $\phi^{(1)}_{\alpha}(\bm q)$ and by projecting to $L=0$ state, 
$\phi_{\alpha}^{(1)}$ is given by 
\begin{align}
\phi_{\alpha}^{(1)}\phi^{4\beta}_{\bm 0}(\bm R)={\cal N}(\zeta)\frac{1}{4\pi}\int d{\bm e}\, \phi_{\alpha}^{(1)}(q\bm e),\ \ \ \ \ \zeta=\frac{1}{8}\beta {\bm q}^2,
\label{def.impurity}
\end{align}
where $\bm e$ is a unit vector,  $|\bm e|=1$, $\int d{\bm e}$ stands for the integration on a sphere with radius 1, 
and ${\cal N}(\zeta)$ is a normalization 
constant to make $\langle \phi^{(1)}_{\alpha}|\phi^{(1)}_{\alpha}\rangle=1$. 
With the use of the overlap 
\begin{align}
\langle \phi^{(1)}_{\alpha}(\bm q) |\phi^{(1)}_{\alpha}(\bm q')\rangle&=\frac{1}{4}\sum_{i,j=1}^4\langle \phi_{\bm q}(i)|\phi_{\bm q'}(j)\rangle \langle \Omega_i|\Omega_j\rangle=\frac{6}{4}
\sum_{i=1}^4\langle \phi_{\bm q}(i)|\phi_{\bm q'}(i)\rangle-\frac{2}{4}\sum_{i\neq j=1}^4\langle \phi_{\bm q}(i)|\phi_{\bm q'}(j)\rangle \notag \\
&=6\big[\langle \phi^{\rm orb}_{\bm q}(423,1)|\phi^{\rm orb}_{\bm q'}(423,1)\rangle-\langle \phi^{\rm orb}_{\bm q}(423,1)|\phi^{\rm orb}_{\bm q'}(143,2)\rangle\big]\notag \\
&=6\, e^{-\frac{3\beta}{16}(\bm q^2+\bm q'^2)}\Big[ {\rm cosh}\big(\frac{3\beta}{8} \bm q\cdot \bm q'\big)-{\rm cosh}\big(\frac{\beta}{8} \bm q\cdot \bm q'\big)\Big],
\end{align}
where $\langle \Omega_i|\Omega_j \rangle=8\delta_{i,j}-2$ is used, 
${\cal N}(\zeta)$ is found to be 
\begin{align}
{\cal N}(\zeta)=\zeta^{\frac{1}{2}}(1-e^{-2\zeta})^{-\frac{3}{2}}.
\end{align}
For $\zeta \to 0$, ${\cal N}(\zeta) \to \frac{1}{2\sqrt{2}\zeta}$ and 
 $\phi_{\alpha}^{(1)}$ approaches a 2$\hbar \omega$ excited configuration  
\begin{align}
\phi_{\alpha}^{(1)}\phi^{4\beta}_{\rm 0}(\bm R) \to \frac{\beta}{3\sqrt{2}}\phi^{\rm orb}_{\alpha}\sum_{i=1}^4 (\bm R-\bm r_i)^2\Omega_i,\ \ \ \ \ \phi^{\rm orb}_{\alpha}=\prod_{i=1}^4 \phi^{\beta}_{\bm 0}({\bm r}_i).
\label{T=1.zerolimit}
\end{align}
Note that the $T=0$ main configuration, $\phi^{(0)}_{\alpha}$, of $\alpha$ particle 
centered at $\bm 0$ reads 
\begin{align}
\phi_{\alpha}^{(0)}\phi^{4\beta}_{\rm 0}(\bm R)=\phi^{\rm orb}_{\alpha}\Omega,\ \ \ \ \ 
\Omega=\frac{1}{\sqrt{2}}\big\{|(00)00,(11)00\rangle-|(11)00,(00)00\rangle\big\}.
\label{S=T=0.alpha}
\end{align}
$\Omega$ is a totally antisymmetric spin-isospin function belonging to the 
representation $[1^4]$ of the symmetric group $S_4$.

To estimate the isospin impurity $\epsilon$ in Eq.~(\ref{alpha*}), 
I simplify $\phi_{\alpha}^{(1)}$, instead of Eq.~(\ref{def.impurity}), by a finite sum of $\phi^{(1)}_{\alpha}(\bm q)$: 
\begin{align}
\phi_{\alpha}^{(1)}\phi^{4\beta}_{\rm 0}(\bm R)={\overline {\cal N}}(\zeta)\frac{1}{3}\big[\phi^{(1)}_{\alpha}(q{\bm e}_x) + \phi_{\alpha}^{(1)}(q\bm e_y)+\phi_{\alpha}^{(1)}(q\bm e_z)\big],
\label{distortedconf}
\end{align}
where, e.g., $\bm e_x$ is a unit vector in $x$ direction, 
and ${\overline {\cal N}}(\zeta)$ is a normalization constant:
\begin{align}
{\overline {\cal N}}(\zeta)=(1+e^{-6\zeta}-e^{-2\zeta}-e^{-4\zeta})^{-\frac{1}{2}}=(1-e^{-2\zeta})^{-1}(1+e^{-2\zeta})^{-\frac{1}{2}}.
\end{align}
${\overline {\cal N}}(\zeta)$ is only slightly smaller than ${\cal N}(\zeta)$ 
for small $\zeta$: E.g., the difference is less than 3\% for $\zeta \leq 0.4$.  

The value $\epsilon$ is determined by solving the secular equation
\begin{align}
\left(
\begin{array}{cc}
H_{00} & H_{01} \\
H_{01} & H_{11} \\
\end{array}
\right)
\left(
\begin{array}{c}
\sqrt{1-\epsilon^2} \\
\epsilon \\
\end{array}
\right)=E
\left(
\begin{array}{c}
\sqrt{1-\epsilon^2} \\
\epsilon \\
\end{array}
\right),
\end{align}
where 
\begin{align}
H_{00}=\langle \phi_{\alpha}^{(0)}|H|\phi_{\alpha}^{(0)} \rangle,\ \ \ \ \ 
H_{11}=\langle \phi_{\alpha}^{(1)}|H|\phi_{\alpha}^{(1)} \rangle,\ \ \ \ \ 
H_{01}=\langle \phi_{\alpha}^{(0)}|H|\phi_{\alpha}^{(1)} \rangle.
\end{align}
The solution of the secular equation is
\begin{align}
E=\frac{1}{2}\Big[H_{00}+H_{11}-\sqrt{(H_{00}-H_{11})^2+4H_{01}^2}\, \Big],\ \ \ \ \ 
\epsilon=-\frac{H_{01}}{|H_{01}|}\frac{1}{\sqrt{1+(\frac{H_{01}}{E-H_{00}})^2}}.
\label{E.epsilon}
\end{align}

The two-nucleon interaction is assumed to be a central potential, $V(r)=e^{-\rho {\bm r}^2}(w+bP^{\sigma}-hP^{\tau}-mP^{\sigma}P^{\tau})$, and the Coulomb potential. 
$H_{00}$ reads 
\begin{align}
H_{00}=\frac{9\hbar^2 \beta}{4m_N}+6(w+m)\big(\frac{\beta}{\beta+2\rho}\big)^{\frac{3}{2}}+\sqrt{\frac{2\beta}{\pi}}e^2.
\end{align}
$H_{01}$ is contributed by the Coulomb potential. 
Let $P^{\rm C}_{12}$ 
denote an operator that acts on two nucleons to select only two-proton 
states, that is, it is unity 
when both the first nucleon and the second nucleon are protons but vanishes 
otherwise. Its isospin matrix element is given by the Clebsch-Gordan coefficient as 
\begin{align}
\langle (T_{12}T_{34})T0 |P^{\rm C}_{12}|(T_{12}' T_{34}')T'0 \rangle=
\delta_{T_{12},1}\delta_{T_{12}',1}\delta_{T_{34},1}\delta_{T_{34}',1}\langle \, 1\, -\!1\, \, 1 \,1\,|T\,0\,\rangle \langle \,1\, -\!1\, \, 1\, 1\,|T'\,0\,\rangle.
\end{align}
This confirms that $\langle \phi^{(0)}_{\alpha}|V_C|\phi^{(1)}_{\alpha}\rangle$ has a non-vanishing  
contribution only through the isospin states between $|(11)00\rangle$ of $\phi^{(0)}_{\alpha}\phi^{4\beta}_{\rm 0}(\bm R)$ and $|(11)10\rangle$ of $\phi^{(1)}_{\alpha}(\bm q)$: 
\begin{align}
&\langle \phi^{(0)}_{\alpha}\phi^{4\beta}_{\rm 0}(\bm R)|V_C|\phi^{(1)}_{\alpha}(\bm q)\rangle
=6\big(\frac{1}{2}\big)\sum_{i=1}^4\langle \phi^{(0)}_{\alpha}\phi^{4\beta}_{\rm 0}(\bm R)|{V_C}_{12}|\phi_{\bm q}(i)\Omega_i \rangle\notag \\
&= -\sqrt{\frac{3}{2}}e^2 
\big\langle \phi^{\rm orb}_{\alpha}\big|\frac{1}{r_{12}}\big|\phi_{\bm q}(1)
+\phi_{\bm q}(2)-\phi_{\bm q}(3)-\phi_{\bm q}(4)\big\rangle 
=-\frac{2\sqrt{6} e^2}{q}e^{-\frac{3\beta}{16} {\bm q}^2}\Big[{\rm erf}\big(\sqrt{\frac{\beta }{8}}q\big)-\sqrt{\frac{\beta}{2\pi}}q\Big],
\end{align}
which leads to 
\begin{align}
H_{01}=-\sqrt{\frac{2\beta}{\pi}}e^2 \sqrt{6}\, {\overline {\cal N}}(\zeta)
e^{-\frac{3}{2}\zeta}h(\sqrt{\zeta}),
\end{align}
where $h(x)=\frac{\sqrt{\pi}}{2}\frac{1}{x}{\rm erf}(x)-1=\sum_{n=1}^{\infty}\frac{(-1)^n}{(2n+1)n!}x^{2n}$. 

Finally I calculate $H_{11}$. 
The contribution of the kinetic energy to $H_{11}$ is obtained from 
\begin{align}
&\langle \phi^{(1)}_{\alpha}(\bm q)|T|\phi^{(1)}_{\alpha}(\bm q')\rangle \notag \\ 
&=4(\frac{1}{2})^2\sum_{i,j=1}^4\langle \phi_{\bm q}(i)|T_1|\phi_{\bm q'}(j)\rangle 
\langle \Omega_i|\Omega_j \rangle 
=6\sum_{i=1}^4 \langle \phi_{\bm q}(i)|T_1|\phi_{\bm q'}(i)\rangle 
-2 \sum_{j>i=1}^4 \big(\langle \phi_{\bm q}(i)|T_1|\phi_{\bm q'}(j)\rangle+
\langle \phi_{\bm q}(j)|T_1|\phi_{\bm q'}(i)\rangle \big)\notag \\
&=6 \Big[\langle \phi_{\bm q}(1)|T_1|\phi_{\bm q'}(1)\rangle+3\langle \phi_{\bm q}(2)|T_1|\phi_{\bm q'}(2)\rangle \Big]
-6 \Big[\langle \phi_{\bm q}(1)|T_1|\phi_{\bm q'}(2)\rangle+\langle \phi_{\bm q}(2)|T_1|\phi_{\bm q'}(3)\rangle + (\bm q \leftrightarrow \bm q')\Big]\notag \\
&=\frac{3\hbar^2 \beta}{4m_N}\Big[e^{-\frac{3\beta}{16}(\bm q-\bm q')^2}\big\{12-\frac{3\beta}{8}(\bm q-\bm q')^2\big\}-e^{-\frac{\beta}{16}(\bm q-\bm q')^2-\frac{\beta}{8}(\bm q+\bm q')^2}\big\{12-\frac{\beta}{16}(\bm q-\bm q')^2-\frac{\beta}{32}(3\bm q+\bm q')^2-\frac{\beta}{32}(\bm q+3\bm q')^2\big\}
\notag \\
&\qquad \quad +(\bm q' \to -\bm q')\Big].
\end{align}
Subtracting the c.m. kinetic energy and projecting to $L=0$, I get 
\begin{align}
\langle \phi_{\alpha}^{(1)}|T|\phi_{\alpha}^{(1)} \rangle=\frac{\hbar^2\beta}{m_N}
\Big[\frac{9}{4}+\frac{1+3e^{-2\zeta}}{e^{2\zeta}-e^{-2\zeta}}\zeta\Big].
\end{align}
The contribution of the potential energy to $H_{11}$ is calculated from
\begin{align}
\langle \phi^{(1)}_{\alpha}(\bm q)|V|\phi^{(1)}_{\alpha}(\bm q')\rangle
&= 6\big(\frac{1}{2}\big)^2 \sum_{i,j=1}^4 \langle \phi_{\bm q}(i)|V_{12}|\phi_{\bm q'}(j)\rangle \notag \\
&\times \big\{w\langle \Omega_i|\Omega_j \rangle +b\langle \Omega_i|P^{\sigma}_{12}|\Omega_j \rangle -h \langle \Omega_i|P^{\tau}_{12}|\Omega_j \rangle
-m \langle \Omega_i|P^{\sigma}_{12}P^{\tau}_{12}|\Omega_j \rangle + 
\langle \Omega_i|P^{\rm C}_{12}|\Omega_j \rangle \big\},
\end{align}
where the spin-isospin matrix elements read as follows:
\begin{align}
&\Big( \langle \Omega_i|P^{\sigma}_{12}|\Omega_j \rangle \Big)=
\left(
\begin{array}{rrrr}
0 & -4 & 2 & 2 \\
-4 & 0 & 2 & 2 \\
2 & 2 & 0 & -4 \\
2 & 2 & -4 & 0 \\
\end{array}
\right),\ \ \ \ \ 
\Big( \langle \Omega_i|P^{\tau}_{12}|\Omega_j \rangle \Big)=
\left(
\begin{array}{rrrr}
4 & 0 & -2 & -2 \\
0 & 4 & -2 & -2 \\
-2 & -2 & 0 & 4 \\
-2 & -2 & 4 & 0 \\
\end{array}
\right),\notag \\
&\Big( \langle \Omega_i|P^{\sigma}_{12}P^{\tau}_{12}|\Omega_j \rangle \Big)=
\left(
\begin{array}{rrrr}
2 & -6 & 2 & 2 \\
-6 & 2 & 2 & 2 \\
2 & 2 & -6 & 2 \\
2 & 2 & 2 & -6 \\
\end{array}
\right),\ \ \ \ \ 
\Big( \langle \Omega_i|P^{\rm C}_{12}|\Omega_j \rangle \Big)=
\left(
\begin{array}{rrrr}
1 & 1 & -1 & -1 \\
1 & 1 & -1 & -1 \\
-1 & -1 & 1 & 1 \\
-1 & -1 & 1 & 1 \\
\end{array}
\right).
\end{align}
The orbital matrix element $O_{ij}=\langle \phi_{\bm q}(i)|V_{12}|\phi_{\bm q'}(j)\rangle$ is specified by five types of integrals as follows: 
\begin{align}
\Big( O_{ij} \Big)=
\left(
\begin{array}{cccc}
O_{11} & O_{12} & O_{13} & O_{13} \\
O_{12} & O_{11} & O_{13} & O_{13} \\
\widetilde{O_{13}} & \widetilde{O_{13}} & O_{33} & O_{34} \\
\widetilde{O_{13}} & \widetilde{O_{13}} & O_{34} & O_{33} \\
\end{array}
\right).
\end{align}
Here, e.g., $\widetilde{O_{13}}$ indicates that its matrix element is obtained by the interchange of $\bm q \leftrightarrow \bm q'$ in $O_{13}$. I obtain 
\begin{align}
\langle \phi^{(1)}_{\alpha}(\bm q)|V_N|\phi^{(1)}_{\alpha}(\bm q')\rangle
= 6\big(\frac{1}{2}\big)^2&\,  \Big[\, (12w-8h-4m)O_{11}+(-4w-8b+12m)O_{12}\notag \\
&+(-8w+8b+8h-8m)O_{13}+(-8w+8b+8h-8m)\widetilde{O_{13}}\notag \\
&+(12w+12m)O_{33}+(-4w-8b-8h-4m)O_{34}\Big],\notag \\
\langle \phi^{(1)}_{\alpha}(\bm q)|V_C|\phi^{(1)}_{\alpha}(\bm q')\rangle
= 6\big(\frac{1}{2}\big)^2&\,  \Big[\, 2O_{11}+2O_{12}
-4O_{13}-4\widetilde{O_{13}}
+2O_{33}+2O_{34}\Big].
\end{align}
Substituting appropriate $O_{ij}$'s for the nuclear potential and the Coulomb potential 
and projecting to $L=0$ leads to the potential energy matrix element as follows:
\begin{align}
\langle \phi^{(1)}_{\alpha}|V|\phi^{(1)}_{\alpha} \rangle
&=[{\overline {\cal N}}(\zeta)]^2\big(\frac{\beta}{\beta+2\rho}\big)^{\frac{3}{2}}
 \Big[(3w-2h-m)(e^{-3\zeta}+e^{-2\eta})^2+(-w-2b+3m)e^{-2\zeta}(e^{-\zeta}+e^{-2\eta})^2\notag \\
&\qquad +(-4w+4b+4h-4m)e^{-2\zeta-\eta}(1+e^{-\zeta})^2+(3w+3m)(1+e^{-3\zeta})^2\notag \\
&\qquad +(-w-2b-2h-m)e^{-2\zeta}(1+e^{-\zeta})^2\Big]+\sqrt{\frac{2\beta }{\pi}}e^2
(1-\Delta_C), 
\end{align}
where $\eta=\frac{2\rho }{\beta+2\rho}\zeta$ and 
\begin{align}
\Delta_C=[{\overline {\cal N}}(\zeta)]^2\Big[2 e^{-2\zeta}(1+e^{-\zeta})^2h(\sqrt{\zeta})-2e^{-3\zeta}h(\sqrt{2\zeta})-\frac{1}{2}(1+e^{-2\zeta})h(\sqrt{4\zeta})\Big].
\end{align}

Two examples are given. In the case of 
Volkov No. 1 potential with $m=0.6$~\cite{volkov65}, a choice of $\beta=0.54$ fm$^{-2}$ and $q=1.8$ fm gives $H_{00}=-27.07182, \, H_{01}=0.22405, \, H_{11}=26.08126$ MeV, respectively, leading to $E=-27.07277$ MeV and $\epsilon=-4.21\times 10^{-3}$. 
Minnesota potential with $u=0.94$~\cite{thompson77}, $\beta=0.55$ fm$^{-2}$ and $q=1.8$ fm gives 
$H_{00}=-24.39133, \, H_{01}=0.22568, \, H_{11}=28.62857$ MeV, respectively, predicting $E=-24.39229$ MeV and $\epsilon=-4.26\times 10^{-3}$. Both 
cases predict almost the same value for $\epsilon$. The $T=1$ impurity component 
of $\alpha$ particle is estimated to be about 0.0018 $\%$, 
in good agreement with 0.0019$\%$ of a realistic calculation using AV18+UX
interaction~\cite{wiringa}. Note, however, that the latter also predicts the $T=2$ component to be 0.0030 $\%$.  
$\epsilon$ varies only a 
little as a function of $q$. In the limit of $q \to 0$, the values of 
$E$ and $\epsilon$ turn out to be  $-27.07277$ MeV and 
$-4.21\times 10^{-3}$ for 
Volkov potential and $-24.39234$ MeV and $-4.13\times 10^{-3}$ for 
Minnesota potential, respectively. This insensitivity is mainly due to the fact that $H_{11}$ is by far larger than $H_{00}$. In what follows I use the $q\to 0$ limit~(\ref{T=1.zerolimit}) as the $T=1$ configuration of $\alpha$ particle.
  
If the change of the normalization of the $N$ $\alpha$-particle wave function is negligible due to the replacement of 
$\phi^{(0)}_{\alpha}$ by $\phi^{(1)}_{\alpha}$,   
the isospin impurities in the present model are  
estimated to be roughly 9$\epsilon^2\approx 0.016$\,\% in $^{12}$C and 
16$\epsilon^2\approx 0.028$\,\% in $^{16}$O, respectively. The isospin-mixing 
rate required to fit 
the $E1$ transition probability  between the $1^-_1$ and $0^+_1$ states of $^{16}$O was on 
the order of a few percents~\cite{pdesc87}, which appears too large in view of 
the density functional estimate~\cite{satula09} that reports the impurity rate  
of less than 1\% even in $^{40}$Ca. Since no IS $E1$ operators are 
taken into account in the calculation of Ref.~\cite{pdesc87}, a full 
calculation including both the IV and IS $E1$ operators may require less 
mixing rate. In fact the ground-state impurities predicted by the isospin-projected
density functional approach with the SLy4 Skyrme force and
Coulomb force being the only source of isospin symmetry violation are
respectively 0.051\% in $^{12}$C and 
0.102\% in $^{16}$O~\cite{satula}. 
Actually the value of $\epsilon$ depends on not only the Coulomb potential but also 
charge-symmetry breaking forces, etc., so that it appears reasonable to set 
$\epsilon$ to fit the $E1$ transition rate in $^{16}$O.

\subsection{Calculation of isovector electric dipole matrix element}
\label{cal.isfE1}

As suggested in Eq.~(\ref{E1mat.ele}), the IV $E1$ matrix element $A_{{\rm IV}}$ 
is obtained through 
\begin{align}
\frac{1}{\epsilon}A_{\rm IV}=\big\langle \phi(\{\bm S\})|\sum_{i=1}^{4N} e\big(\frac{1}{2}-t_3(i)\big){\cal Y}_{1\mu}(\bm r_i)|\phi'(\{\bm S'\})\big\rangle
+\big\langle \phi'(\{\bm S\})|\sum_{i=1}^{4N} e\big(\frac{1}{2}-t_3(i)\big){\cal Y}_{1\mu}(\bm r_i)|
\phi(\{\bm S'\})\big\rangle,
\label{isfE1}
\end{align}
where, e.g.,  $\phi'(\{\bm S'\})$ denotes the $T=1$ configuration comprising  
$N$ components, each of which is defined by replacing   
one of $\phi^{(0)}_{\alpha}$'s in $\phi(\{\bm S'\})$  by the 2$\hbar \omega$ excited configuration of Eq.~(\ref{T=1.zerolimit}), $\phi^{(1)}_{\alpha}$, (cf. Eq.~(\ref{Nalpha.WF})):
\begin{align}
\phi'(\{\bm S'\})
=\frac{1}{\sqrt{4!}^N}{\cal A}_{4N} \Bigg\{ \sum_{l=1}^N
\phi^{4\beta}_{\bm S'_l}(\bm R_l)\phi^{(1)}_{\alpha}(l) 
\Big(\prod_{j=1\atop j \neq l}^{N}\phi^{4\beta}_{\bm S'_j}(\bm R_j)\phi^{(0)}_{\alpha}(j) \Big)\Bigg\}.
\label{Nalpha.T=1.impurity}
\end{align} 
Let  $M_{1\mu}(\bm S, \bm S')$ denote the first term on the right-hand side of Eq.~(\ref{isfE1}). 
The second term turns out to be given by $(-1)^{\mu} (M_{1-\mu}(\bm S', \bm S))^*$. 
I focus on the calculation of $M_{1\mu}(\bm S, \bm S')$ in what follows.

The spin-isospin symmetry of $\phi(\{\bm S\})$ is characterized by the direct 
product of $N\, \Omega$'s, i.e.,  $[1^4]\cdots [1^4]$ symmetry of $S_4$.  
This special property makes it possible to 
simplify the calculation of $M_{1\mu}(\bm S, \bm S')$. 
Let ${\bm r}_{l_i}$ with $l_i=l+(i-1)N\ (i=1,\ldots,4)$ denote the single-particle coordinates of the $l$th $\alpha$ particle.  Its c.m. coordinate is $\bm R_l=\frac{1}{4}\sum_{i=1}^4 \bm r_{l_i}$. See Eq.~(\ref{partial.antsymmetrizer}). Let $d_{1\mu}(l)$ denote the IV $E1$ operator acting on 
the $l$th $\alpha$ particle:
\begin{align}
d_{1\mu}(l)=e \sum_{i=1}^4 \big(\frac{1}{2}-t_3(l_i)\big){\cal Y}_{1\mu}(\bm r_{l_i}).
\end{align}
Acting the $E1({\rm IV})$ operator on $\phi'(\{\bm S'\})$ leads to two different types of 
configurations:  
\begin{align}
&\sum_{i=1}^{4N} e\big(\frac{1}{2}-t_3(i)\big){\cal Y}_{1\mu}(\bm r_i)|\phi'(\{\bm S'\})\rangle\notag \\
&=\frac{1}{\sqrt{4!}^N}{\cal A}_{4N}\Big\{ 
\sum_{l,l'=1 \atop l \neq l'}^N \phi^{4\beta}_{\bm S'_1}(\bm R_1)\phi^{(0)}_{\alpha}(1)\cdots d_{1\mu}(l') \phi^{4\beta}_{\bm S'_{l'}}(\bm R_{l'})\phi^{(0)}_{\alpha}(l') \cdots 
\phi^{4\beta}_{\bm S'_l}(\bm R_l)\phi^{(1)}_{\alpha}(l)\cdots \phi^{4\beta}_{\bm S'_N}(\bm R_N)\phi^{(0)}_{\alpha}(N)\Big\} \notag \\
&+\frac{1}{\sqrt{4!}^N}{\cal A}_{4N}\Big\{\sum_{l=1}^N \phi^{4\beta}_{\bm S'_1}(\bm R_1)\phi^{(0)}_{\alpha}(1)\cdots
d_{1\mu}(l)\phi^{4\beta}_{\bm S'_l}(\bm R_l)\phi^{(1)}_{\alpha}(l)\cdots \phi^{4\beta}_{\bm S'_N}(\bm R_N)\phi^{(0)}_{\alpha}(N)\Big\}.
\label{E1xT=1}
\end{align}
In the first type $d_{1\mu}$ acts on the $\alpha$-particle wave function with 
$T=0$, while in the second type it acts on the excited configuration with $T=1$. 
A non-vanishing contribution to $M_{1\mu}(\bm S, \bm S')$ occurs 
only when either of the acted configurations has the same 
spin-isospin symmetry as $\phi(\{\bm S\})$, i.e., $\Omega^N$. From this criterion, apparently any configurations of the first type have no 
contribution to $M_{1\mu}(\bm S, \bm S')$.  

The configuration of the second type is characterized by 
$d_{1\mu}(l)\phi^{4\beta}_{\bm S'_l}(\bm R_l)\phi^{(1)}_{\alpha}(l)$, which reads as 
\begin{align}
d_{1\mu}(l)\phi^{4\beta}_{\bm S'_l}(\bm R_l)\phi^{(1)}_{\alpha}(l)={\cal O}_{1\mu}(l)
\prod_{i=1}^4\phi_{\bm S'_l}^{\beta}(\bm r_{l_i}),\ \ \ \ \ {\cal O}_{1\mu}(l) 
=\frac{\beta}{3\sqrt{2}}d_{1\mu}(l)\sum_{j=1}^4(\bm R_l-\bm r_{l_j})^2\Omega_{l_j}.
\label{second.type}
\end{align}
Here,  Eq.~(\ref{T=1.zerolimit}) is used for $\phi^{4\beta}_{\bm S'_l}(\bm R_l)\phi^{(1)}_{\alpha}(l)$. 
In order for this wave function to contribute 
to $M_{1\mu}(\bm S, \bm S')$,  the spin-isospin part of 
${\cal O}_{1\mu}(l)$ must contain the component of $\Omega$, which  is 
nothing but $\langle \Omega | {\cal O}_{1\mu}(l)\rangle \Omega$. 
Note that $\Omega_{l_1}, \ldots,  \Omega_{l_4}$, and $\Omega$ are defined in Eqs.~(\ref{impurity.state}) and~(\ref{S=T=0.alpha}), 
respectively, with the nucleon labels 1, 2, 3, 4 being replaced by $l,\, l+N,\, l+2N, \, l+3N$. 
The contribution of Eq.~(\ref{second.type}) is then determined by 
$\langle \Omega | {\cal O}_{1\mu}(l)\rangle \phi^{4\beta}_{\bm S'_l}(\bm R_l)\phi^{(0)}_{\alpha}(l)$. The quantity $\langle \Omega | {\cal O}_{1\mu}(l)\rangle$ is 
calculated as follows: 
\begin{align}
\langle \Omega | {\cal O}_{1\mu}(l)\rangle=
\frac{\beta}{3\sqrt{2}} \sum_{j=1}^4 \langle \Omega |d_{1\mu}(l)| \Omega_{l_j}\rangle ({\bm R}_l-\bm r_{l_j})^2.
\label{omega.comp}
\end{align}
Using $\langle \Omega \big|\frac{1}{2}-t_3(l_i)\big|\Omega_{l_j} \rangle=\frac{1}{\sqrt{6}}(1-4\delta_{i,j})$ leads to 
$\langle \Omega |d_{1\mu}(l)| \Omega_{l_j}\rangle
=\frac{4}{\sqrt{6}}e{\cal Y}_{1\mu}(\bm R_l-\bm r_{l_j})$. It follows that 
$\langle \Omega | {\cal O}_{1\mu}(l)\rangle=\frac{2}{3\sqrt{3}} {\cal D}_{1\mu}(l)$,
where ${\cal D}_{1\mu}(l)$ is  a kind of translation-invariant $E1$ operator acting on $\phi^{(0)}_{\alpha}(l)$: 
\begin{align}
{\cal D}_{1\mu}(l)=e \beta \sum_{j=1}^4(\bm R_l-{\bm r}_{l_j})^2
{\cal Y}_{1\mu}(\bm R_l-{\bm r}_{l_j}).
\label{isfE1.op}
\end{align}
After all, 
$M_{1\mu}(\bm S, \bm S')$ reduces to the matrix element of the IS $E1$ type, 
\begin{align}
M_{1\mu}(\bm S, \bm S')=\frac{2}{3\sqrt{3}}\sum_{l=1}^N \big\langle \phi(\{\bm S\})|
\frac{1}{\sqrt{4!}^N}{\cal A}_{4N}\Big\{ 
{\cal D}_{1\mu}(l) 
\prod_{k=1}^{N} \Big( \phi^{4\beta}_{\bm S'_k}(\bm R_k)\phi^{(0)}_{\alpha}(k)\Big) \Big\}\big\rangle.
\label{isfE1'}
\end{align}

A direct use of ${\cal D}_{1\mu}(l)$ in Eq.~(\ref{isfE1'}) leads to a fairly involved calculation. To avoid it I make use of an integral representation of  ${\cal D}_{1\mu}(l)$ (see Eqs.~(\ref{f.g.relation}) and~(\ref{exp.spherical})):
\begin{align}
{\cal D}_{1\mu}(l)=e \frac{5}{4\pi \sqrt{\beta}}\sum_{j=1}^4\int d{\bm e}\, Y_{1\mu}(\bm e) \frac{d^3}{d\alpha^3}\Big|_{\alpha=0} e^{\sqrt{\beta}\alpha \bm e\cdot (\bm R_l-\bm r_{l_j})},
\end{align}
where $\bm e$ is a unit vector.  The action of ${\cal D}_{1\mu}(l)$ on the $l$th $\alpha$-particle wave function reads  
\begin{align}
{\cal D}_{1\mu}(l)\prod_{i=1}^4 \phi^{\beta}_{\bm S'_l}(\bm r_{l_i})
=e\frac{5}{4\pi \sqrt{\beta}}\int d{\bm e}\, Y_{1\mu}(\bm e) \frac{d^3}{d\alpha^3}\Big|_{\alpha=0}\, 
e^{\frac{3}{8}\alpha^2} \sum_{j=1}^4 \prod_{i=1}^4\phi^{\beta}_{\bm S'_l+\frac{\epsilon^{(ij)}}{\sqrt{\beta}}\alpha \bm e}(\bm r_{l_i}),
\label{D.wf}
\end{align}
where $\epsilon^{(ij)}=\frac{1}{4}-\delta_{i,j}$.  Because of 
$\sum_{i=1}^4 \epsilon^{(ij)}=0$,  the c.m. position 
of the $l$th $\alpha$-particle remains at $\bm S'_l$ independent of $\alpha \bm e$. More exactly, one of the four nucleons is centered at 
$\bm S'_l-\frac{3}{4\sqrt{\beta}}\alpha \bm e$ and the other three nucleons 
are at $\bm S'_l+\frac{1}{4\sqrt{\beta}}\alpha \bm e$. 
The use of Eq.~(\ref{D.wf}) in Eq.~(\ref{isfE1'}) enables one to get $M_{1\mu}(\bm S, \bm S')$ through the overlap of the Slater determinants:
\begin{align}
M_{1\mu}(\bm S, \bm S')=e\frac{2}{3\sqrt{3}}\frac{5}{4\pi \sqrt{\beta}}\int d{\bm e}\, Y_{1\mu}(\bm e) \frac{d^3}{d\alpha^3}\Big|_{\alpha=0} \, 
e^{\frac{3}{8}\alpha^2} \sum_{l=1}^N \langle \phi(\{\bm S\})|\phi_l(\bm S')\rangle,
\label{isf.E1.mat.ele}
\end{align}
where $\phi_l(\bm S')$ is defined by replacing $\prod_{i=1}^4\phi^{\beta}_{\bm S'_l}(\bm r_{l_i})\chi_i$ in $\phi(\bm S')$ (see Eq.~(\ref{Nalpha.SD'})) by $\sum_{j=1}^4 \prod_{i=1}^4\phi^{\beta}_{\bm S'_l+\frac{\epsilon^{(ij)}}{\sqrt{\beta}}\alpha \bm e}(\bm r_{l_i})\chi_i$. Let  $b_l^{(ij)}$ denote an $N\times N$ matrix defined by replacing $b_{kl}$ of the matrix $b$ by 
\begin{align}
\langle \phi^{\beta}_{\bm S_k}|\phi^{\beta}_{\bm S'_l+\frac{\epsilon^{(ij)}}{\sqrt{\beta}}\alpha \bm e}\rangle,\ \ \ \ \ (k=1,\ldots, N).
\end{align}
The sum of the overlap of the Slater determinants in Eq.~(\ref{isf.E1.mat.ele}) reduces to   
\begin{align}
&\sum_{l=1}^N \langle \phi(\{\bm S\})|\phi_l(\bm S')\rangle=\sum_{l=1}^N \sum_{j=1}^4 
\prod_{i=1}^4 {\rm det}\, b_l^{(ij)}=4 \sum_{l=1}^N \, {\rm det}\, b_l^{(11)} 
({\rm det}\, b_l^{(21)})^3\notag \\
&=4 \sum_{l=1}^N 
\left(\sum_p \epsilon(p)b_{p_11}\cdots e^{-\frac{\beta}{4}(\bm S_{p_l}-\bm S'_l+\frac{3}{4\sqrt{\beta}}\alpha \bm e)^2}\cdots b_{p_N N}\right)
  \left(\sum_q \epsilon(q)b_{q_11}\cdots e^{-\frac{\beta}{4}(\bm S_{q_l}-\bm S'_l-\frac{1}{4\sqrt{\beta}}\alpha \bm e)^2}\cdots b_{q_N N} \right)^3\notag \\
&=4 \, e^{-\frac{3}{16}\alpha^2}\sum_{l=1}^N \left(\sum_p \epsilon(p) e^{-\frac{\beta}{4}\sum_{i=1}^N (\bm S_{p_i}-\bm S'_i)^2-\frac{3\sqrt{\beta}}{8}\alpha \bm e\cdot (\bm S_{p_l}-\bm S'_l)}\right)
  \left(\sum_q \epsilon(q)e^{-\frac{\beta}{4}\sum_{i=1}^N (\bm S_{q_i}-\bm S'_i)^2+\frac{\sqrt{\beta}}{8}\alpha \bm e\cdot (\bm S_{q_l}-\bm S'_l) } \right)^3,
\label{E1.ovlap}
\end{align}
where the cubic power consists of at most 2600 terms for $N=4$~\cite{count}. 
The resulting  $\alpha \bm e$-dependent terms  take the form 
\begin{align}
e^{-\frac{\sqrt{\beta}}{8}\alpha \bm e\cdot (3(\bm S_{p_l}-\bm S'_l)- (\bm S_{q_l}-\bm S'_l)- (\bm S_{q'_l}-\bm S'_l)- (\bm S_{q''_l}-\bm S'_l)   )} = e^{-\frac{\sqrt{\beta}}{8}\alpha \bm e\cdot {\bm D}_l},
\end{align}
where 
\begin{align}
\bm D_l=3(\bm S_{p_l}-\bm S'_l)- (\bm S_{q_l}-\bm S'_l)- (\bm S_{q'_l}-\bm S'_l)- (\bm S_{q''_l}-\bm S'_l)=3\bm S_{p_l}-\bm S_{q_l}-\bm S_{q'_l}-\bm S_{q''_l}.
\end{align}
Here, $q,\, q',\, q''$ are permutation labels used to expand the cubic term. 
The integral in Eq.~(\ref{isf.E1.mat.ele}) is found to be 
\begin{align}
e\frac{2}{3\sqrt{3}}\frac{5}{4\pi \sqrt{\beta}}\int d{\bm e}\, Y_{1\mu}(\bm e) \frac{d^3}{d\alpha^3}\Big|_{\alpha=0}\,e^{\frac{3}{8}\alpha^2}\,  4\, e^{-\frac{3}{16}\alpha^2} e^{-\frac{\sqrt{\beta}}{8}\alpha \bm e\cdot {\bm D}_l}
=-e\frac{1}{3\sqrt{3}}\Big[\frac{15}{8}{\cal Y}_{1\mu}(\bm D_l)+\frac{\beta}{64}{\bm D_l}^2{\cal Y}_{1\mu}(\bm D_l)\Big].
\label{alp.e.op}
\end{align} 

Combining Eqs.~(\ref{isf.E1.mat.ele}),~(\ref{E1.ovlap}), and~(\ref{alp.e.op}) and 
taking the sum for $l,\, p,\, q,\, q',\, q''$ gives the matrix element 
$M_{1\mu}(\bm S, \bm S')$. 
The intrinsic matrix element $M_{1\mu}(\bm S, \bm S')_{\overline{\bm S}=\overline{\bm S'}=\bm 0}$ is obtained by replacing the sum of type,  $\sum_{i=1}^N (\bm S_{p_i}-\bm S'_i)^2$, 
in Eq.~(\ref{E1.ovlap}) by $\sum_{i=1}^N (\bm S_{p_i}-\bm S'_i)^2-N(\overline{\bm S}-\overline{\bm S'})^2$.

\section{Matrix elements between correlated Gaussians}
\label{matele.cg}

The CG basis $f_{KLM}(u,A,\bm x)$ of Eq.~(\ref{def.fLM}) has proven to 
provide an accurate solution for a Schr\"odinger equation of a 
few-body system. Refer to \cite{varga98,usukura98,nemura02,horiuchi07,horiuchi08,horiuchi12,horiuchi14} for a variety of 
applications and to \cite{mitroy13} for a review article. The advantage of 
the CG basis is that one can calculate the matrix elements of most operators 
analytically, which enables one to obtain a precise solution by superposing 
the CGs with different parameters. The systems 
treated in Refs.~\cite{varga98,usukura98,nemura02,horiuchi07,horiuchi08,horiuchi12,horiuchi14}  contain only few identical particles, so that 
fermion/boson symmetry requirements can easily be incorporated with only few permutations.  
In the present case, however, $f_{KLM}(u,A,\bm x)$ is tied with $\phi^{\rm in}(N\alpha)$, the product of 
the internal wave functions of $\alpha$ particles. This 
leads to two problems: One is that an enormous number of permutations, $(4N)!$, 
appears. The other is that an analytic evaluation of a matrix element is 
impractical because of so many terms. A way to get out from this difficulty is 
to make use of the generating function of $f_{KLM}(u,A,\bm x)$~\cite{varga95,book98}.

\subsection{Generating correlated Gaussians}
\label{generatingCG}
The CG basis~(\ref{CGbasis}) is generated from the Slater 
determinant of the GWPs~(\ref{Nalpha.SD}).  To show this, I 
introduce the inter-cluster relative and total c.m. coordinates for 
both the physical coordinate and the generator (or parameter) coordinate: 
\begin{align}
&\bm x_j=\frac{1}{j}\sum_{i=1}^j\bm R_i-\bm R_{j+1},\ \ \ \ \ \bm s_j=\frac{1}{j}\sum_{i=1}^j\bm S_i-\bm S_{j+1}\ \ \ (j=1,\ldots,N-1),\notag \\ 
&\bm R=\frac{1}{N}\sum_{i=1}^N \bm R_i=\frac{1}{4N}\sum_{i=1}^{4N}\bm r_i,\ \ \ \ \ \overline{\bm S}=\frac{1}{N}\sum_{i=1}^N \bm S_i.
\end{align}
They are compactly expressed in terms of an $N\times N$ transformation 
matrix $U$ as follows:
\begin{align}
\left(\begin{array}{c}
\bm x_1 \\
\vdots \\
\bm x_{N-1} \\
\bm R \\
\end{array}
\right)
=U
\left(\begin{array}{c}
\bm R_1 \\
\vdots \\
\bm R_{N-1} \\
\bm R_N \\
\end{array}
\right),\ \ \ \ \ 
\left(\begin{array}{c}
\bm s_1 \\
\vdots \\
\bm s_{N-1} \\
\overline{\bm S} \\
\end{array}
\right)
=U
\left(\begin{array}{c}
\bm S_1 \\
\vdots \\
\bm S_{N-1} \\
\bm S_N \\
\end{array}
\right).
\label{coord.transf}
\end{align}
$U$ and its inverse $U^{-1}$ are explicitly given by
\begin{align}
U=\left(
\begin{array}{ccccc}
1           & -1          & 0  & \cdots & 0 \\
\frac{1}{2} & \frac{1}{2} & -1 & \cdots & 0 \\
\vdots      & \vdots     &    &  & \vdots \\
\frac{1}{N-1} & \frac{1}{N-1} & \cdots & \cdots & -1 \\
\frac{1}{N} & \frac{1}{N} & \cdots & \cdots & \frac{1}{N} \\
\end{array}
\right),\ \ \ \ \ 
U^{-1}=\left(
\begin{array}{ccccc}
\frac{1}{2} & \frac{1}{3} & \cdots  & \frac{1}{N} & 1 \\
-\frac{1}{2} & \frac{1}{3} & \cdots & \frac{1}{N} & 1 \\
0      & -\frac{2}{3}   &    & \vdots & \vdots \\
\vdots & \vdots &  & \vdots & \vdots \\
0 & 0 & \cdots & -\frac{N-1}{N} & 1 \\
\end{array}
\right).
\end{align}
In the case of $N=4$, $\bm x_1$ and $\bm x_2$ are two Jacobi coordinates to 
describe 3 $\alpha$-particle motion in $^{12}$C and 
$\bm x_3$ is the relative distance vector between $^{12}$C and $\alpha$ particle.

The transformation of the coordinates makes it possible to 
rewrite the product of the GWPs to 
\begin{align}
\prod_{i=1}^N \phi_{\bm S_i}^{4\beta}(\bm R_i)\phi_{\alpha}^{(0)}(i)=\phi^{\rm in}(N\alpha)
\prod_{i=1}^N \phi_{\bm S_i}^{4\beta}(\bm R_i)=
\phi^{\rm in}(N\alpha)\phi_{\overline{\bm S}}^{4N\beta}(\bm R)\prod_{i=1}^{N-1}\phi_{\bm s_i}^{4\nu_i \beta}(\bm x_i),
\end{align}
where $\nu_i=\frac{i}{i+1}$ and $\phi^{\rm in}(N\alpha)$ is defined in Eq.~(\ref{prod.alpha.internal}). By letting $\bm x$ and $\bm s$ stand for the set $\{\bm x_1,\ldots,\bm x_{N-1}\}$ and $\{\bm s_1,\ldots,\bm s_{N-1}\}$, respectively, the product of the GWPs for the $N\alpha$-particle relative motion  reads as 
\begin{align}
\prod_{i=1}^{N-1}\phi_{\bm s_i}^{4\nu_i \beta}(\bm x_i)=\left(\frac{4\beta}{\pi}\right)^{\frac{3}{4}(N-1)}e^{-\frac{1}{2}\tilde{\bm x}\Gamma \bm x+\tilde{\bm x}\Gamma \bm s-\frac{1}{2}\tilde{\bm s}\Gamma \bm s},
\label{prod.GWP}
\end{align}
where $\Gamma=(4 \beta \nu_i\delta_{i,j})$ is an $(N-1)\times (N-1)$ diagonal matrix, and its determinant is ${\rm det}\, \Gamma=\frac{1}{N}(4\beta)^{N-1}$. 
By leaving out the total c.m. function $\phi_{\overline{\bm S}}^{4N\beta}(\bm R)$ 
from $\phi(\{\bm S\})$, 
the intrinsic wave function of $N\alpha$-particle system takes the form 
\begin{align}
\phi^{\rm in}(\{\bm s\})\equiv \phi^{\rm in}(\{\bm S\}) =\frac{1}{{\sqrt{4!}}^N} \left(\frac{4\beta}{\pi}\right)^{\frac{3}{4}(N-1)}
{\cal A}_{4N}\left\{ e^{-\frac{1}{2}\tilde{\bm x}\Gamma \bm x+\tilde{\bm x}\Gamma \bm s-\frac{1}{2}\tilde{\bm s}\Gamma \bm s}\phi^{\rm in}(N\alpha) \right\}.
\label{gcm.wf.nalpha}
\end{align}

To prove that $\phi^{\rm in}(\{\bm s\})$ serves to generate $\Psi_{K}^{LM}(u,A)$ of Eq.~(\ref{CGbasis}), I express $f_{KLM}(u,A,\bm x)$ by an integral transform of $ e^{-\frac{1}{2}\tilde{\bm x}\Gamma \bm x+\tilde{\bm x}\Gamma \bm s-\frac{1}{2}\tilde{\bm s}\Gamma \bm s}$, as detailed in Eq.~(6.33) of Ref.~\cite{book98}.
First I note that $f_{KLM}(u,A,\bm x)$ is generated as follows~\cite{varga95}:
\begin{align}
f_{KLM}(u,A,\bm x)= \frac{1}{B_{KL}}\int d\bm e \, Y_{LM}(\bm e)\frac{d^{2K+L}}
{d\alpha^{2K+L}}\Big|_{\alpha=0}g(\alpha \bm e u, A, \bm x)
\label{f.g.relation}
\end{align}
with 
\begin{align}
g(\alpha \bm e u, A, \bm x)=e^{-\frac{1}{2}\tilde{\bm x}A\bm x+\alpha \bm e \tilde{u}\bm x}, 
\ \ \ \ \ B_{KL}=\frac{4\pi (2K+L)!}{(2K)!!(2K+2L+1)!!},
\label{gene.fn}
\end{align}
where $\bm e$ is a unit vector. A key relation here is 
\begin{align}
(\bm e\cdot \bm v)^n=\sum_{k, \lambda \atop 2k+\lambda=n} \sum_{m=-\lambda}^{\lambda} B_{k \lambda}Y_{\lambda m}^*(\bm e)
\bm v^{2k}{\cal Y}_{\lambda m}(\bm v)
=\sum_{k, \lambda \atop 2k+\lambda=n}B_{k \lambda}(-1)^{\lambda}\sqrt{2\lambda+1}{\bm v}^{2k}[Y_{\lambda}(\bm e)\times {\cal Y}_{\lambda}(\bm v)]_{00}, 
\label{exp.spherical}
\end{align}
valid for an arbitrary vector $\bm v$. The function $e^{-\frac{1}{2}\tilde{\bm x}\Gamma \bm x+\tilde{\bm x}\Gamma \bm s-\frac{1}{2}\tilde{\bm s}\Gamma \bm s}$ in Eq.~(\ref{gcm.wf.nalpha}) 
has the form of Eq.~(\ref{gene.fn}):
\begin{align}
e^{-\frac{1}{2}\tilde{\bm x}\Gamma \bm x+\tilde{\bm x}\Gamma \bm s-\frac{1}{2}\tilde{\bm s}\Gamma \bm s}=g(\Gamma \bm s,\Gamma, \bm x)e^{-\frac{1}{2} \tilde{\bm s}\Gamma \bm s}=g(\Gamma \bm x,\Gamma, \bm s)e^{-\frac{1}{2}\tilde{\bm x}\Gamma \bm x}.
\label{exponent}
\end{align}
For an integral kernel, $Ce^{-\frac{1}{2}\tilde{\bm s}Q\bm s+\tilde{\bm v}\bm s}=Cg(\bm v, Q, \bm s)$, where 
$\bm v$ is a column vector of $N-1$ dimension and $Q$ is an $(N-1)\times (N-1)$ symmetric matrix, 
an integral transform of Eq.~(\ref{exponent}) is given by 
\begin{align}
C\int d{\bm s} \, e^{-\frac{1}{2}\tilde{\bm s}Q\bm s+\tilde{\bm v}\bm s}
e^{-\frac{1}{2}\tilde{\bm x}\Gamma \bm x+\tilde{\bm x}\Gamma \bm s-\frac{1}{2}\tilde{\bm s}\Gamma \bm s}
=C\left(\frac{(2\pi)^{N-1}}{{\rm det}(Q+ \Gamma)}\right)^{\frac{3}{2}}
e^{-\frac{1}{2} \tilde{\bm x}\Gamma \bm x+\frac{1}{2}\tilde{\bm w}(Q+ \Gamma)^{-1}\bm w},
\label{int.g&g}
\end{align}
where $d\bm s=d\bm s_1\ldots d\bm s_{N-1}$ and $\bm w=\bm v + \Gamma \bm x$.
The integral again takes the form of $g$ as expected~\cite{book98}. 
The quadratic term of $\bm w$ in Eq.~(\ref{int.g&g}) is 
\begin{align}
\tilde{\bm w}(Q+ \Gamma)^{-1}\bm w
= \tilde{\bm x}\Gamma (Q+ \Gamma)^{-1} \Gamma \bm x+2 \tilde{\bm v}(Q+ \Gamma)^{-1}\Gamma \bm x+\tilde{\bm v}(Q+ \Gamma)^{-1}\bm v.
\end{align}
The integral~(\ref{int.g&g}) reduces to $g(\alpha\bm eu, A, \bm x)$, 
provided that the following conditions are satisfied:
\begin{align}
 \Gamma-\Gamma (Q+\Gamma)^{-1}\Gamma=A, \ \ \ \ \ 
 \Gamma (Q+ \Gamma)^{-1} \bm v=\alpha \bm e u,\ \ \ \ \   
C\left(\frac{4\beta}{\pi}\right)^{\frac{3}{4}(N-1)}
\left(\frac{(2\pi)^{N-1}}{{\rm det}(Q+ \Gamma)}\right)^{\frac{3}{2}}
e^{\frac{1}{2}\tilde{\bm v}(Q+ \Gamma)^{-1}\bm v}=1.
\end{align}
The unknowns $Q, \bm v$, and $C$ are found to be 
\begin{align}
&Q=A (\Gamma-A)^{-1}\Gamma=A+A(\Gamma -A)^{-1}A,\ \ \ \ \  \bm v=\alpha \bm e \Gamma(\Gamma -A)^{-1}{u},\notag \\
&C=\frac{1}{N^3}\left(\frac{16\beta^3}{\pi}\right)^{\frac{3}{4}(N-1)}[{\rm det}(\Gamma-A)]^{-\frac{3}{2}}e^{-\frac{1}{2}\alpha^2\tilde{u}(\Gamma -A)^{-1}u}.
\label{def.Q.v}
\end{align} 

To summarize, the CG basis for $N\alpha$-particle 
system is generated from the  multi-cluster wave function of the generator coordinate method type through the 
manipulation defined below~\cite{varga95}:
\begin{align}
\Psi_{K}^{LM}(u,A)&=\frac{1}{N^3}\left(\frac{16\beta^3}{\pi}\right)^{\frac{3}{4}(N-1)}[{\rm det}(\Gamma-A)]^{-\frac{3}{2}}\notag \\
&\times \frac{1}{B_{KL}} \int d\bm e \, Y_{LM}(\bm e)
\frac{d^{2K+L}}{d\alpha^{2K+L}}\Big|_{\alpha=0}e^{-\frac{1}{2}\alpha^2\tilde{u}(\Gamma -A)^{-1}u} 
\int d{\bm s} \, g(\bm v,Q,\bm s) \phi^{\rm in}(\{\bm s\}).
\label{cg.sld}
\end{align}
Especially $\Psi^{00}_0(u,A)$ takes a simple form, and $u$ turns out to be a 
redundant parameter:
\begin{align}
\Psi^{00}_0(u,A)=\frac{1}{N^3}\left(\frac{16\beta^3}{\pi}\right)^{\frac{3}{4}(N-1)}[{\rm det}(\Gamma-A)]^{-\frac{3}{2}}\frac{1}{\sqrt{4\pi}}\int d{\bm s}\, e^{-\frac{1}{2}\tilde{\bm s}Q{\bm s}} \phi^{\rm in}(\{\bm s\}).
\end{align}

\subsection{Transformation of matrix elements from Slater determinants to correlated Gaussians}
\label{transf.slater.cg}

Applying Eq.~(\ref{cg.sld}) one gets the matrix element of a spherical tensor operator ${\cal O}_{\kappa \mu}$ between the CGs from the one between the Slater determinants as follows:
\begin{align}
\langle \Psi_{K}^{LM}&(u,A)|{\cal O}_{\kappa \mu}|\Psi_{K'}^{L'M'}(u',A')\rangle=\frac{1}{N^6}\left(\frac{16\beta^3}{\pi}\right)^{\frac{3}{2}(N-1)}[{\rm det}(\Gamma-A)\, {\rm det}(\Gamma - A')]^{-\frac{3}{2}} 
 \frac{1}{B_{KL} B_{K'L'}}\notag \\
&\times \iint d{\bm e}\, d{\bm e}'\, Y_{LM}^*(\bm e)\, Y_{L'M'}(\bm e')\,
 \frac{d^{2K+L+2K'+L'}}{d\alpha^{2K+L}d\alpha'^{2K'+L'}}\Big|_{\alpha=0,\alpha'=0} \, 
 e^{-\frac{1}{2}\alpha^2\tilde{u}(\Gamma -A)^{-1}u-\frac{1}{2}\alpha'^2\tilde{u'}(\Gamma -A')^{-1}u'}
\notag \\
&\times \iint d{\bm s}\, d{\bm s'}\, g(\bm v, Q,\bm s)\, g(\bm v', Q', \bm s')\, 
\langle \phi^{\rm in}(\{\bm s\})|{\cal O}_{\kappa \mu}|\phi^{\rm in}(\{\bm s'\})\rangle,
\label{grand.formula}
\end{align}
where $\bm v$ and $Q$  are defined in Eq.~(\ref{def.Q.v}), and 
$\bm v'$ and $Q'$ are defined by $A'$ and $u'$ similarly. 
Equation~(\ref{grand.formula}) consists of three steps: (1) the integration for $\bm s$ 
and $\bm s'$, (2) the differentiation with respect to $\alpha$ and 
$\alpha'$ followed by setting $\alpha=\alpha'=0$, and (3) the integration for $\bm e$ and $\bm e'$. 
The step (1) 
is specific to the microscopic formulation. The steps (2) and (3) are familiar routines in the CG basis, as used in 
many examples~\cite{varga95,book98,book03,suzuki08,aoyama12}. 
   
I start from the step (1). Removing the c.m. coordinates $\overline{\bm S}$ and $\overline{\bm S'}$, I rename $\bm s_1,\ldots, \bm s_{N-1}, \bm s_1',\ldots, \bm s_{N-1}'$ to 
$\bm t$. As shown in Secs.~\ref{Nalpha.me.cm.free} and~\ref{isf.alpha}, 
$\langle \phi^{\rm in}(\{\bm s\})|{\cal O}_{\kappa \mu}|\phi^{\rm in}(\{\bm s'\})\rangle$ of interest consists of a number of terms of the form 
\begin{align}
P(\bm t)e^{-\frac{1}{2}\tilde{\bm t}W\bm t},
\end{align}
where $P(\bm t)$ is a polynomial of $\bm t$ and $W$ is a $(2N-2)\times (2N-2)$ 
symmetric matrix. The product of $g(\bm v, Q,\bm s)g(\bm v', Q', \bm s')$ in Eq.~(\ref{grand.formula}) is compactly written as
$g(\bm v, Q,\bm s)g(\bm v', Q', \bm s')=e^{-\frac{1}{2}\tilde{\bm t}{\cal Q}\bm t
+\tilde{\bm v}\bm t}$ 
with 
\begin{align}
{\cal Q}=
\left(
\begin{array}{cc}
Q & 0\\
0 & Q'\\
\end{array}
\right),\ \ \ \ \ 
\bm v=
\left(
\begin{array}{c}
\alpha \bm e \Gamma(\Gamma-A)^{-1}u \\
\alpha' \bm e' \Gamma(\Gamma-A')^{-1}u'\\
\end{array}
\right)
\equiv
\left(
\begin{array}{c}
\alpha \bm e w \\
\alpha' \bm e' w'\\
\end{array}
\right),
\label{int.formula}
\end{align}
where both $w$ and $w'$ 
are column vectors of $N-1$ dimension. 
The step (1) is completed by performing the integration 
\begin{align}
\iint d{\bm s}\, d{\bm s'}g(\bm v, Q,\bm s)g(\bm v', Q', \bm s')
\langle \phi^{\rm in}(\{\bm s\})|{\cal O}_{\kappa \mu}|\phi^{\rm in}(\{\bm s'\})\rangle = \sum 
\int d{\bm t} \, P(\bm t)\, e^{-\frac{1}{2}\tilde{\bm t}Z\bm t +\tilde{\bm v}\bm t},
\end{align}
where $d{\bm t}=d{\bm s} d{\bm s}'=d{\bm t}_1\cdots d{\bm t}_{2N-2}$ and $Z=W+{\cal Q}$. 
Table~\ref{tab.integral} lists  integration formulas for typical $P(\bm t)$. 
See also, e.g., Exercise 6.2 and Table 7.1 of Ref.~\cite{book98}.
As an example, I show the case of $P(\bm t)=\tilde{\bm t}Y\bm t$: 
\begin{align}
I=\frac{d}{d\lambda}\Big|_{\lambda=0} \int d{\bm t}\, e^{-\frac{1}{2}\tilde{\bm t}Z\bm t +\lambda \tilde{\bm t}Y\bm t + \tilde{\bm v}\bm t}
=\frac{d}{d\lambda}\Big|_{\lambda=0} I_1(Z-2\lambda Y).
\end{align}
See Table~\ref{tab.integral} for $I_1(Z)$. The operation $\frac{d}{d\lambda}\big|_{\lambda=0}$ can be done by using 
the well-known formulas of matrices, 
\begin{align}
(A+\lambda B)^{-1}=A^{-1}-\lambda A^{-1}BA^{-1}+\cdots,\ \ \ \ \ 
{\rm det}(A+\lambda B)={\rm det}A+\lambda ({\rm Tr}A^{-1}B) \, {\rm det}A+\cdots,
\end{align} 
where $\cdots$ stands for higher-order terms in $\lambda$, leading to 
\begin{align}
I_1(Z-2\lambda Y)= \Big[1+\lambda \Big(3{\rm Tr}Z^{-1}Y+
\tilde{\bm v}Z^{-1}YZ^{-1}\bm v\Big) +\cdots \Big]I_1(Z), 
\end{align}
the use of which readily confirms the formula in Table~\ref{tab.integral}. 
Similarly the formulas of other cases are obtained from those of known formulas together with some additional expansions, e.g., 
\begin{align}
&{\cal Y}_{1\mu}(\tilde {\zeta} (Z-2\lambda Y)^{-1}\bm v)={\cal Y}_{1\mu}(\tilde{\zeta} Z^{-1}\bm v)+2\lambda {\cal Y}_{1\mu}(\tilde{\zeta} Z^{-1}YZ^{-1}\bm v)+\cdots,\notag \\
&(Z-2\lambda W)^{-1}Y(Z-2\lambda W)^{-1}=Z^{-1}YZ^{-1}+2\lambda Z^{-1}(YZ^{-1}W+WZ^{-1}Y)Z^{-1}+\cdots.
\end{align}
   
\begin{table*}[h]
\caption{Integral formulas for $I=\int d{\bm t} \, P(\bm t)\, e^{-\frac{1}{2}\tilde{\bm t}Z\bm t +\tilde{\bm v}\bm t}$, where $\bm t=(\bm t_i)$ is a column vector of 
$2N-2$ dimension and $d\bm t$ stands for $d\bm t_1 \ldots d\bm t_{2N-2}$. $\bm v=(\bm v_i)$ is a column vector of $2N-2$ dimension, and $Z$ is a $(2N-2)\times (2N-2)$ symmetric matrix. $I$ is expressed as $I=F(\bm v)I_1(Z)$, where $I_1(Z)=\Big(\frac{(2\pi)^{2N-2}}{{\rm det} Z}\Big)^{\frac{3}{2}} e^{\frac{1}{2}\tilde{\bm v}Z^{-1}\bm v}$. 
To specify a polynomial $P(\bm t)$, a column vector $\zeta=(\zeta_i)$ of $2N-2$ 
dimension and $(2N-2)\times (2N-2)$ symmetric matrices, $Y$ and $V$, are used.   }
\begin{tabular}{lccl}
\hline\hline
 $P(\bm t)$ &&& $F(\bm v)$ \\
\hline\hline
1 &&& 1 \\
$\tilde{\bm t}Y\bm t$ &&& $3{\rm Tr}Z^{-1}Y+\tilde{\bm v}Z^{-1}YZ^{-1}\bm v$ \\
${\cal Y}_{2\mu}(\tilde{\zeta}\bm t)$ &&& ${\cal Y}_{2\mu}(\tilde{\zeta}Z^{-1}\bm v)$\\
${\cal Y}_{1\mu}(\tilde{\zeta}\bm t)$ &&& ${\cal Y}_{1\mu}(\tilde{\zeta}Z^{-1}\bm v)$\\
$(\tilde{\bm t}Y\bm t) {\cal Y}_{1\mu}(\tilde{\zeta}\bm t)$ &&& $\big(3{\rm Tr}Z^{-1}Y+\tilde{\bm v}Z^{-1}YZ^{-1}\bm v\big){\cal Y}_{1\mu}(\tilde{\zeta}Z^{-1}\bm v)+2{\cal Y}_{1\mu}(\tilde{\zeta}Z^{-1}YZ^{-1}\bm v)$\\ 
$(\tilde{\bm t}V\bm t)( \tilde{\bm t}Y\bm t) {\cal Y}_{1\mu}(\tilde{\zeta}\bm t)$ &&& $\big[
6{\rm Tr}\,Z^{-1}YZ^{-1}V+\big(3{\rm Tr}Z^{-1}Y+\tilde{\bm v}Z^{-1}YZ^{-1}\bm v\big)\big(3{\rm Tr}Z^{-1}V+\tilde{\bm v}Z^{-1}VZ^{-1}\bm v\big)$\\
  &&& $   +2\tilde{\bm v}Z^{-1}(YZ^{-1}V + VZ^{-1}Y)Z^{-1}\bm v \big]{\cal Y}_{1\mu}(\tilde{\zeta}Z^{-1}\bm v)$\\
  &&& $+2\big(3{\rm Tr}Z^{-1}Y+\tilde{\bm v}Z^{-1}YZ^{-1}\bm v\big){\cal Y}_{1\mu}(\tilde{\zeta}Z^{-1}VZ^{-1}\bm v)
   +2\big(3{\rm Tr}Z^{-1}V+\tilde{\bm v}Z^{-1}VZ^{-1}\bm v\big){\cal Y}_{1\mu}(\tilde{\zeta}Z^{-1}YZ^{-1}\bm v)$\\
  &&& $+4{\cal Y}_{1\mu}(\tilde{\zeta}Z^{-1}(YZ^{-1}V+VZ^{-1}Y)Z^{-1}\bm v)$\\
${\cal Y}_{3\mu}(\tilde{\zeta}\bm t)$ &&&  ${\cal Y}_{3\mu}(\tilde{\zeta}Z^{-1}\bm v)$\\
\hline\hline
\end{tabular}
\label{tab.integral}
\end{table*}

To complete the step (2), one has to make explicit $\alpha$- and $\alpha'$-dependence of the integral  $I$. Their dependence comes from 
$\bm v$ as defined in Eq.~(\ref{int.formula}). Table~\ref{tab.integral} shows that   
three kinds of $\bm v$-dependent terms appear: 
$\tilde{\bm v}Z_1 \bm v$, 
${\cal Y}_{1\mu}(\tilde{\zeta}Z_2 \bm v)$, and ${\cal Y}_{2\mu}(\tilde{\zeta}Z_2 \bm v)$, 
where both $Z_1$ and $Z_2$ are $(2N-2)\times (2N-2)$ symmetric matrices.
By defining a column vector $\omega$ of $2N-2$ dimension by 
$\omega_i=w_i, \, \omega_{N-1+i}=w'_i$ for $i=1,\ldots, N-1$, it follows that 
\begin{align}
&\tilde{\bm v}Z_1\bm v=p\alpha^2+p'\alpha'^2+q\alpha \alpha'\bm e\cdot \bm e',\ \ \ \ \ 
{\cal Y}_{1\mu}(\tilde{\zeta}Z_2\bm v)=\gamma \alpha Y_{1\mu}(\bm e)+\gamma' 
\alpha' Y_{1\mu}(\bm e'),\notag \\
&{\cal Y}_{2\mu}(\tilde{\zeta}Z_2\bm v)={\cal Y}_{2\mu}(\gamma \alpha \bm e +\gamma' \alpha' \bm e')=\gamma^2 \alpha^2 Y_{2\mu}(\bm e)+\gamma'^2 \alpha'^2 Y_{2\mu}(\bm e')+\sqrt{\frac{40\pi}{3}}\gamma \gamma' \alpha \alpha' [Y_1(\bm e)\times Y_1(\bm e')]_{2\mu},
\end{align}
where
\begin{align}
&p=\sum_{i,j=1}^{N-1}\omega_i(Z_1)_{ij}\omega_j,\ \ \ \ \ 
p'=\sum_{i,j=N}^{2N-2}\omega_i(Z_1)_{ij}\omega_j,\ \ \ \ \ 
q=2\sum_{i=1}^{N-1}\sum_{j=N}^{2N-2}\omega_i(Z_1)_{ij}\omega_j,\notag \\
&\gamma=\sum_{i=1}^{2N-2}\, \sum_{j=1}^{N-1}{\zeta}_i(Z_2)_{ij}\omega_j,\ \ \ \ \ 
\gamma'=\sum_{i=1}^{2N-2}\, \sum_{j=N}^{2N-2}{\zeta}_i(Z_2)_{ij}\omega_j.
\end{align}
The general form of the integral $I$ including the exponential 
$e^{-\frac{1}{2}\alpha^2\tilde{u}(\Gamma -A)^{-1}u-\frac{1}{2}\alpha'^2\tilde{u'}(\Gamma -A')^{-1}u'}$ is summarized as follows: 
\begin{align}
&{\rm for\ overlap\ and\ central\ potential\ of\ Gaussian\ form\ factor}\ (\kappa=0)\notag\\
&\qquad \quad  C_0\,e^{\rho \alpha^2+\rho'\alpha'^2+ \rho_1 \alpha \alpha'\bm e\cdot \bm e'}, 
\label{general.overlap}\\
&{\rm for\  kinetic\ energy\ and\ squared\ radius}\ (\kappa=0)\notag \\
&\qquad \quad \big(C_0+C\alpha^2+C'\alpha'^2+C_1\alpha \alpha' \bm e\cdot \bm e'\big)\,e^{\rho \alpha^2+\rho'\alpha'^2+ \rho_1 \alpha \alpha'\bm e\cdot \bm e'},\\
&{\rm for\ }E2\ {\rm operator}\ (\kappa=2)\notag \\
&\qquad \quad \big(Q\alpha^2 Y_{2\mu}(\bm e)+Q'\alpha'^2 Y_{2\mu}(\bm e')+Q_1 \alpha \alpha' 
[Y_1(\bm e)\times Y_1(\bm e')]_{2\mu}\big)\,e^{\rho \alpha^2+\rho'\alpha'^2+ \rho_1 \alpha \alpha'\bm e\cdot \bm e'},\label{general.E2.op}\\ 
&{\rm for\ } E1\ {\rm operator}\ (\kappa=1)\notag \\
&\qquad \quad \Big( \sum_{n+n'=0,1,2}C_{n,n'}\alpha^{2n}\alpha'^{2n'} +\sum_{n+n'=0,1}C'_{n,n'}\alpha^{2n+1} \alpha'^{2n'+1} \bm e\cdot \bm e'+ C''\alpha^2\alpha'^2(\bm e\cdot \bm e')^2
 \Big)\notag \\
&\qquad \quad \times (D\alpha Y_{1\mu}(\bm e)+D'\alpha' Y_{1\mu}(\bm e'))\,e^{\rho \alpha^2+\rho'\alpha'^2+ \rho_1 \alpha \alpha'\bm e\cdot \bm e'},
\label{general.E1.op}\\
&{\rm for\ }E3\ {\rm operator}\ (\kappa=3)\notag \\
&\qquad \quad \big(O\alpha^3 Y_{3\mu}(\bm e)+O'\alpha'^3 Y_{3\mu}(\bm e')+O_1 \alpha^2 \alpha' 
[Y_2(\bm e)\times Y_1(\bm e')]_{3\mu}+ O'_1 \alpha \alpha'^2 
[Y_1(\bm e)\times Y_2(\bm e')]_{3\mu} \big)\notag \\
&\qquad \quad \times e^{\rho \alpha^2+\rho'\alpha'^2+ \rho_1 \alpha \alpha'\bm e\cdot \bm e'}.\label{general.E3.op}
\end{align}
Note that $\rho,\, \rho',\, \rho_1$ as well as the coefficients $C$, etc., 
depend on the respective operators. 
The step (2) is completed by expanding the exponential function $e^{\rho \alpha^2+\rho'\alpha'^2+ \rho_1 \alpha \alpha'\bm e\cdot \bm e'}$ in power series,
\begin{align}
e^{\rho \alpha^2+\rho'\alpha'^2+ \rho_1 \alpha \alpha'\bm e\cdot \bm e'}
=\sum_{k=0}^{\infty}\sum_{k'=0}^{\infty}\sum_{l=0}^{\infty}\frac{\rho^k \rho'^{k'} \rho_1^l}{k! k'! l!}\alpha^{2k+l}\alpha'^{2k'+l}(\bm e\cdot \bm e')^l, 
\end{align}
and by collecting all those terms that have powers of $\alpha^{2K+L} \alpha'^{2K'+L'}$ in order for 
$\frac{d^{2K+L+2K'+L'}}{d\alpha^{2K+L}d{\alpha'}^{2K'+L'}}\Big|_{\alpha=0,\alpha'=0}$ to give 
nonzero contribution. Since $K, L$ and $K', L'$ are 0 or small positive 
integers, 
actually few terms contribute to the matrix element, 
$\langle \Psi_{K}^{LM}(u,A)|{\cal O}_{\kappa \mu}|\Psi_{K'}^{L'M'}(u',A')\rangle$. 

The step (3) is the angular momentum projection prescribed in 
Eq.~(\ref{grand.formula}). After the step (2) all the surviving scalar terms 
contain $(\bm e\cdot \bm e')^n$, where $n$'s are non-negative integers. 
They are multiplied by $[Y_l(\bm e)\times Y_{l'}(\bm e')]_{\kappa \mu}$, 
where $\kappa $ is 0, 1, or 2. See Eqs.~(\ref{general.overlap}) to~(\ref{general.E1.op}). 
Using the formula~(\ref{exp.spherical}), 
one can easily perform the angular integration:
\begin{align}
&\iint d{\bm e}\, d{\bm e}'\, Y_{LM}^*(\bm e)\, Y_{L'M'}(\bm e')\, 
[Y_l(\bm e)\times Y_{l'}(\bm e')]_{\kappa \mu}\, 
[Y_{\lambda}(\bm e)\times Y_{\lambda}(\bm e')]_{00}\notag \\
&=(-1)^{l+l'+L}
\sqrt{\frac{2\kappa+1}{(2l+1)(2\lambda+1)}} \, 
C(l \,\lambda \,; L) \, C(l' \, \lambda \,;L')\, 
U(\kappa \,l' \,L \,\lambda \,;l\,L') \,\langle \, L'\,M'\, \kappa \,\mu \,|L\,M \,\rangle ,
\end{align} 
where $U$ is an $R(3)$ (unitary) Racah coefficient~\cite{book98} and 
\begin{align}
C(l\, l'\, ;L)=\sqrt{\frac{(2l+1)(2l'+1)}{4\pi(2L+1)}}\langle \, l\,0\,l'\, 0\,|L\,0\,\rangle.
\end{align}

The general form for the $E1$ operator, Eq.~(\ref{general.E1.op}), indicates 
that  the $E1$ matrix element between 
$L=0$ and $L'=1$ basis states, $\langle \Psi_{K}^{00}(u,A)|E1|\Psi_{K'}^{1M'}(u',A')\rangle$, has a contribution only  from 
$C_{0,0}D'$ if both $K$ and $K'$ are restricted to 0. With the increase of 
$K$ and $K'$, higher-order terms in $\alpha$ and $\alpha'$ contribute to the 
$E1$ matrix element. Whether or not such CGs need to be included in the 
basis set depends on the extent to which they contribute to gaining energy.

\subsection{Correlated Gaussian parameters and reduced $\alpha$-width amplitude}
\label{choice.cgparam}

The quality of the CG description of the states of $^{16}$O depends on how 
well the CG parameters are chosen. A guide to choose 
$A$ and $u$ in particular is found in many examples~\cite{varga98,usukura98,nemura02,horiuchi07,horiuchi08,horiuchi12,horiuchi14}. 
The aim here is to briefly mention its point on the assumption that  
one computes the $^{12}$C($\alpha, \gamma$)$^{16}$O radiative-capture cross 
section in the $R$-matrix method~\cite{desc10}. 

Equation~(\ref{coord.transf}) shows that the relative distance vector, 
$\bm R_i-\bm R_j$, is related to $\bm x$ by 
$\bm R_i-\bm R_j=\sum_{k=1}^{N-1}(U^{-1}_{\,ik}-U^{-1}_{\,jk})\bm x_k\equiv \sum_{k=1}^{N-1}w^{(ij)}_k\bm x_k=\widetilde{w^{(ij)}}\bm x$. Introducing variational 
parameters $d_{ij}$'s, one may parametrize $A$ as~\cite{book98} 
\begin{align}
A=\sum_{j>i=1}^N\frac{1}{d_{ij}^2}\,w^{(ij)}\widetilde{w^{(ij)}},
\label{A.parameter}
\end{align}
which is equivalent to assuming $\tilde{\bm x}A\bm x=\sum_{j>i=1}^N\frac{1}{d_{ij}^2}(\bm R_i-\bm R_j)^2$. Thus $d_{ij}$'s control the distances among the $\alpha$ particles. 

As noted in Sec.~\ref{condition}, first one has to determine 
the ground state of $^{12}$C. The parameter $K$ in Eq.~(\ref{CGbasis}) 
can be set to 0. Its important role 
 is to improve the short-range behavior of the wave function. See, e.g., 
Refs.~\cite{suzuki98,nemura02}.  If the 
interaction between the $\alpha$ particles turns out to be strongly repulsive 
at short distances, 
the use of nonzero $K$ values can be effective. However, no such thing is 
expected to occur in the present case. With $K=0$, $u$ becomes redundant. 
The ground state of $^{12}$C is therefore assumed to take the form (see Eq.~(\ref{CGbasis})) 
\begin{align}
\Psi^{0_1^+0}(^{12}{\rm C})=\frac{1}{\sqrt{4!}^3}{\cal A}_{12}\big\{\Phi_{00}(\bm x)\, \phi^{\rm in}(3\alpha)\big\},
\label{carbon.state}
\end{align}
where the relative motion 
function $\Phi_{00}(\bm x)$ among 3 $\alpha$-particles becomes a combination of Gaussians (see Eq.~(\ref{def.fLM})):
\begin{align}
\Phi_{00}(\bm x)=\sum_i C_i\, \frac{1}{\sqrt{4\pi}}\,e^{-\frac{1}{2}\tilde{\bm x}A_i \bm x},
\end{align}
where $\bm x=\Big(\begin{array}{c}\bm x_1 \\ \bm x_2 \\ \end{array}\Big)$ 
and $A_i$ is a $2\times 2$ symmetric and positive-definite matrix. 
$A_i$ contains only 3 free parameters, and they 
may be parametrized as in Eq.~(\ref{A.parameter}). $C_i$'s are 
determined by solving a Schr\"odinger equation for $^{12}$C. 

In the $R$-matrix method the configuration space is divided into two regions, 
internal and external. In the external region, channel wave 
functions are represented by $^{12}$C$+\alpha$ two-cluster configurations. 
Since the energy region of interest is around the Gamow window, one may assume 
that only the elastic channel is included in calculations. By letting $L$ 
denote the orbital angular momentum between $^{12}$C and $\alpha$ particle,  
the wave function of the elastic channel $\Psi^{LM}_{el}$ with 
parity $\pi=(-1)^L$ takes the form
\begin{align}
\Psi^{LM}_{el}
=\frac{1}{\sqrt{4!}^4}{\cal A}_{16}
\big\{\Phi_{00}(\bm x)\, \chi(x_3){\cal Y}_{LM}(\bm x_3)\, \phi^{\rm in}(4\alpha)\big\},
\label{channel.wf.C+alpha}
\end{align} 
where $\chi(x_3)$ is the radial function for the relative motion 
to be determined. Depending on whether the  energy of the relative motion between $^{12}$C and $\alpha$ particle is 
positive or negative, $\chi(x_3)$ should be smoothly connected to the 
Coulomb wave function or the Whittaker 
function beyond the channel radius~\cite{desc10}. For $x_3$ less than 
the channel radius, 
$\chi(x_3)$ is expanded as $\chi(x_3)=\sum_n c_n {x_3}^{2K_n}e^{-\frac{1}{2}a_n \bm x_3^2}$. $\Psi^{LM}_{el}$ is then written as a combination of the CGs, Eq.~(\ref{CGbasis}), 
\begin{align}
\Psi^{LM}_{el} 
=\sum_{i,n}C_ic_n \frac{1}{\sqrt{4\pi}} \Psi^{LM}_{K_n}(u,A_{i,n}),
\label{external.wf}
\end{align}
where $f_{K_n LM}(u, A_{i,n},\bm x)$ that defines $\Psi^{LM}_{K_n}(u,A_{i,n})$ is characterized by $u$ and $A_{i,n}$ of decoupled form
\begin{align}
u=\left( \begin{array}{c}
0\\
0\\
1\\
\end{array}\right),\ \ \ \ \ 
A_{i,n}=\left(\begin{array}{cc}
A_{i} &  0 \\
0    & a_n \\
\end{array}\right).
\end{align}
Note that $A_i$'s are already known. $K_n=0$ is usually a good choice to get accurate solutions. It is evident that the 
description of the elastic channel wave functions demands no double angular-momentum projection. 

One has to determine the bound states, $0^+_1, 0^+_2, 1^-_1, 2^+_1$, of $^{16}$O.  
The wave function for the respective state $\Psi^{LM}$ is expressed in terms of the CG basis 
functions. See Eq.~(\ref{totalwf}). In the basis set, the configurations $\Psi^{LM}_{K_n}(u,A_{i,n})$ 
of $^{12}$C+$\alpha$ type are included. When $^{12}$C and $\alpha$ particle are 
close enough to interact strongly, however, there is no reason that $^{12}$C 
stays at 
the ground state. To take into account its possibility of distortion, 
one should include the CG basis functions with unrestricted $u$ and $A$ 
parameters. In the latter basis functions the angular momenta  of 
3 $\alpha$-clusters and the 3$\alpha$-$\alpha$ relative motion in general 
have no sharp values~\cite{suzuki98}. 
The number of independent CG parameters is 2 for $u$ and 6 for $A$ in general. 
These numbers are not very large but probably too large to include all basis 
functions that are provided by discretizing $u$ and $A$ in a grid. 
It is practically important to get a high-quality solution with a 
limited number of parameter sets. 
There are number of cases that the stochastic optimization performs quite well, 
leading to a precise solution. Refer to~\cite{varga95,book98,varga94} for its 
basic idea and to~\cite{varga98,usukura98,nemura02,horiuchi07,horiuchi08,horiuchi12,horiuchi14} for several examples.
 
The $R$-matrix formalism requires the channel wave function at the 
channel radius where the inter-cluster antisymmetrization can be neglected. 
To get the asymptotic behavior of the channel wave function, it is useful to express 
the full antisymmetrizer ${\cal A}_{16}$ in terms of the double coset 
decomposition of the permutation group~\cite{hall59},
\begin{align}
{\cal A}_{16}=\frac{1}{\sqrt{16!}} {\cal A}_{12}{\cal A}_4 \Big(1+\sum_m g_m\Big){\cal A}_{12}{\cal A}_4,
\end{align}
where 1 in the round parenthesis corresponds to no exchange of the nucleons 
intervening between $^{12}$C and $\alpha$ particle, and other double coset 
generators $g_m$ induce the intercluster exchange of the nucleons. The number 
of $g_m$'s including 1 is $\frac{16!}{12!4!}=1820$. Beyond the 
channel radius, $\Psi^{LM}_{el}$ of Eq.~(\ref{channel.wf.C+alpha}) can be 
replaced by
\begin{align}
\Psi^{LM}_{el} \to \binom {16}{12}^{-\frac{1}{2}}\Psi^{0^+_10}(^{12}{\rm C})\,\phi_{\alpha}^{(0)}\chi(x_3){\cal Y}_{LM}(\bm x_3).
\end{align}

As mentioned in Introduction, the reduced $\alpha$-width of the subthreshold 
state is a very important quantity 
to determine the radiative-capture cross section. 
It is also called an $\alpha$-particle spectroscopic amplitude. The reduced $\alpha$-width of 
the state $\Psi^{LM}$ to the elastic channel is defined by 
\begin{align}
y_L(r)=\sqrt{\binom {16}{12}}\Big\langle \Psi^{0^+_10}(^{12}{\rm C})\phi^{(0)}_{\alpha}Y_{LM}(\widehat{\bm x_3})\frac{\delta(x_3-r)}{x_3r}\Big|\Psi^{LM}\Big\rangle.
\end{align}
A convenient way to calculate $y_L(r)$ is to approximate the Dirac 
$\delta$-function as 
\begin{align}
\frac{\delta(x_3-r)}{x_3r} \approx \sum_{\nu}f_{\nu}(r)f_{\nu}(x_3),
\end{align}
where $\{f_{\nu}\}$ is a square-integrable `pseudo-complete' set with the 
property $\langle f_{\nu}|f_{\nu'}\rangle=\delta_{\nu, \nu'}$. 
Such set $\{f_{\nu}(r)\}$ may be constructed from a number of functions, 
$r^Le^{-\frac{1}{2}a{\bm r}^2}$, with suitably chosen $a$'s~\cite{ogawa00}. 
The calculation of $y_L(r)$ then reduces to that of the 
overlap, $\big\langle \Psi^{0^+_10}(^{12}{\rm C})\phi^{(0)}_{\alpha}e^{-\frac{1}{2}a{\bm x_3^2}}{\cal Y}_{LM}(\bm x_3)|\Psi^{LM}\big\rangle$. The 
choice of $a$'s depends on how far in $r$ and in what interval of $r$ one wants 
to calculate $y_L(r)$. A practical way of generating $a$'s is to use 
a geometric progression~\cite{hiyama03}.  Examples of applying 
the CGs to calculate $y_L(r)$ are found  in Refs.~\cite{horiuchi08,horiuchi14}.

\section{Summary}
\label{summary}
The $^{12}$C($\alpha, \gamma$)$^{16}$O radiative-capture process near the Gamow 
window proceeds from the $^{12}$C$+\alpha$ continuum to the ground state of $^{16}$O via the electric dipole and electric quadrupole transitions. No resonances are 
present in the energy region of interest, but the $1^-$ and $2^+$ subthreshold states play an important role 
through their reduced $\alpha$-width amplitudes. In addition, the electric 
dipole transition belongs to a class of isospin-forbidden transitions. 
A quantitative prediction of the radiative-capture cross section demands 
due care of these points. 

I have attempted to provide all the needed materials within a fully microscopic 
4 $\alpha$-particle model. The wave function of $\alpha$ 
particle is extended from the $(0s)^4$ configuration to include the 2$\hbar \omega$ excited configuration with $T=1$. Both isovector and isoscalar dipole transitions are taken into account. 
The motion among $\alpha$ particles is described in the correlated Gaussian 
functions. The two-nucleon interaction is assumed to be a central force 
plus the Coulomb potential. All the necessary matrix elements are presented. 
The advantage of the correlated Gaussians is that they can describe 
a variety of four-body structure in both the internal region and the 
external region.  
Although everything appears clear, persistent numerical works for the nuclear 
structure part will be required to reach the goal of 
predicting the $^{12}$C($\alpha, \gamma$)$^{16}$O radiative-capture cross 
section.

\acknowledgments

The author is deeply indebted to D. Baye  for several communications on 
the electric dipole operator for isospin-forbidden transitions. He is grateful 
to N. Itagaki for discussions on the electric dipole transition in $^{12}$C. 
He also thanks W. Satu{\l}a and 
R.~B. Wiringa for providing him with the isospin impurity rates in $^{12}$C 
and $^{16}$O as well as in $\alpha$  particle.


\begin{thebibliography}{99}
\bibitem{count} The number is obtained by counting how many different terms 
appear from $(a_1+\cdots+a_n)^4$, where $n=4!$ and $a_i$'s are assumed to 
differ from each other. The number of terms of type $a_i^4,\, a_i^3a_j,\, a_i^2a_j^2, \, a_i^2a_ja_k$, or $a_ia_ja_ka_l$ is $n,\, 2{}_nC_2,\, {}_nC_2,\, n{}_{n-1}C_2,\, {}_nC_4$, respectively. The total number of terms amounts to $\frac{1}{24}n(n+1)(n+2)(n+3)$, which is 17550. Similarly $(a_1+\cdots+a_n)^3$ contains 2600 terms.
\end{thebibliography}
\end{document}